\documentclass[preprint,showpacs,preprintnumbers,amsmath,amssymb, superscriptaddress,longbibliography,nofootinbib]{revtex4-1}
\usepackage{graphicx} 
\usepackage{xcolor}
\usepackage{amsmath}
\usepackage{hyperref}
\usepackage[top = 2cm, bottom = 2cm, right = 2cm, left =2cm]{geometry}

\usepackage{orcidlink}
\usepackage{float}
\usepackage{hyperref}
\usepackage{amsmath, amsthm, amsfonts,amssymb}
\usepackage[svgnames]{xcolor}
\usepackage{mathtools}
\usepackage{caption}
\usepackage{subcaption}
\usepackage{soul}
\usepackage{appendix}
\usepackage{orcidlink}
\usepackage{booktabs}

\usepackage{tikz}
\usetikzlibrary{decorations.pathmorphing,calc}
\begin{document}

\title{Regular fluid of strings black hole with non trivial core and asymptotic structure by gravitational decoupling}

\author{Milko Estrada \orcidlink{0000-0002-8982-4000}}
\email{milko.estrada@gmail.com}
\affiliation{Departamento de F\'{\i}sica, Facultad de Ciencias, Universidad de Tarapac\'{a}, Casilla 7-D, Arica, Chile}

\author{Luis C. N. Santos \orcidlink{0000-0002-6129-1820}}
\email{luis.santos@ufsc.br}
\affiliation{Departamento de F\'{\i}sica, CFM -- Universidade Federal de Santa Catarina; C.P. 476, CEP 88.040-900, Florian\'{o}polis, SC, Brazil}

\author{Leonardo G. Barbosa \orcidlink{0009-0007-3468-3718}}
\email{leonardo.barbosa@posgrad.ufsc.br}
\affiliation{Departamento de F\'{\i}sica, CFM -- Universidade Federal de Santa Catarina; C.P. 476, CEP 88.040-900, Florian\'{o}polis, SC, Brazil}


\begin{abstract}
Cloud-of-strings (CS) geometries provide an effective description of one-dimensional string distributions. However, their central singularity cannot be removed through the standard  regular black holes (RBH) mechanism based on an effective mass function, since the string sector contributes independently to the ultraviolet structure of the spacetime. Motivated by this observation, we investigate whether string-supported black holes can be consistently regularized under phenomenological assumptions while preserving the CS asymptotics and admitting a physically meaningful string-fluid interpretation. Using the gravitational decoupling method and a phenomenologically constructed screening function, we construct a RBH supported by an effective anisotropic string fluid. We show that the string sector modifies the local spatial curvature while preserving the leading de Sitter behavior near the center, and introduces a longer-range correction dominating the usual Hayward/LQG term. The geometry admits non-extremal and extremal RBH, as well as a regular horizonless compact object. Moreover, the string parameter modifies the temperature and heat-capacity profiles, while the extremal radius remains independent of it at fixed regularization scale. Finally, the scalar quasinormal-mode spectrum exhibits systematic changes in both the oscillation frequencies and damping rates. These results indicate that regularizing string-supported black holes is a physically distinct problem, with the matter sector {affecting} the ultraviolet structure, thermodynamics, and dynamical response of the spacetime.
\end{abstract}

\maketitle

\section{Introduction}\label{sec:Introduction}

The detection of gravitational waves from black hole mergers \cite{LIGOScientific:2016aoc} and the observations of the shadows of supermassive black holes have established these objects as one of the foremost laboratories for testing gravity in the strong-field regime. Nevertheless, most classical black hole solutions predict the existence of a central singularity, where curvature invariants diverge and the geometric description of spacetime breaks down. This problem has motivated the development of a broad class of regular black holes (RBHs), characterized by replacing the singularity with a regular central region, usually associated with a de Sitter core. In this context, identifying physically well-motivated matter sources capable of supporting regular geometries remains an open problem. Among BHs, one of the best-known models is the Hayward solution \cite{Hayward:2005gi}. In this scenario, the energy of a collapsing star condenses into a Planck-density core, and gravitational collapse enters a quantum-gravitational phase associated with Loop Quantum Gravity (LQG), known as a \textit{Planck star} \cite{DeLorenzo:2014pta}, where gravitational attraction is balanced by quantum pressure. As a result, the spacetime develops a repulsive de Sitter core at short distances (see Eq.~\eqref{dSCore}), while at large distances quantum-gravitational effects become almost negligible, giving rise to the characteristic LQG correction $V\sim r^{-4}$ to the gravitational potential. Although subleading with respect to the Newtonian contribution $V\sim r^{-1}$, these corrections could become more accessible at intermediate distances where the Newtonian potential is still dominant but the quantum contribution is not yet completely suppressed.

Along these lines, LQG-inspired geometrical effects have
been investigated in cosmological spacetimes through loop
quantum cosmology (LQC), where classical symmetries are
imposed before quantization using methods inspired by LQG \cite{Ashtekar:2011ni}.
Analogous approaches have also been applied to black-hole
spacetimes \cite{Kelly:2020uwj}. In particular,
quantum-gravitational effects may modify black-hole collapse
and potentially prevent the formation of classical
singularities \cite{Husain:2006cx}. When the density of
collapsing matter approaches the Planck scale, such effects
may lead to a nonsingular transition, as proposed in
Planck-star scenarios \cite{Rovelli:2014cta}.
Although these models remain theoretical, possible
observational signatures have been discussed in the
context of radio astronomy \cite{Rovelli:2017zoa}.
Interestingly, an $r^{-4}$ correction also emerges
in the quantum Oppenheimer--Snyder model
\cite{Lewandowski:2022zce}, where effective LQC corrections
to gravitational collapse yield a modified Schwarzschild
geometry. This provides further motivation for the
LQG-inspired phenomenological framework considered here,
although our screening function is not derived from
the underlying quantum-gravitational dynamics.

Motivated by the search for physically meaningful matter sources capable of supporting black holes geometries, cloud-of-strings distributions \cite{Letelier:1979ej} have attracted considerable attention, see Refs.~\cite{Zafar:2025sxl,Javed:2026tdz,Gogoi:2025ied,Ahmed:2025ojg} for recent examples. In this framework, a macroscopic ensemble of one-dimensional strings is described by a cloud of strings whose energy density scales as $\rho\propto r^{-2}$, modifying the spacetime geometry and giving rise to black hole solutions while preserving spherical symmetry. Subsequently, this formalism was generalized to the case of a fluid of strings by introducing an effective pressure \cite{Letelier:1983du}, allowing for a broader class of matter distributions. More recently, the string-fluid model has been described through a linear equation of state \cite{NunesdosSantos:2025alw}, which reproduces the standard cloud-of-strings behavior in an appropriate limit. In addition, Ref.~\cite{Estrada:2026kea} showed that the interior of the Reissner--Nordström black hole can be modeled by a string fluid. Other recent applications of string fluids can be found in Refs.~\cite{Huo:2026enq,Errehymy:2026ntw}. Thus, constructing a RBH that simultaneously preserves the characteristic cloud-of-strings asymptotics and admits a consistent string-fluid interpretation remains a nontrivial problem. Indeed, unlike vacuum geometries, the cloud-of-strings contribution constitutes an additional gravitational sector that is independent of the mass parameter, suggesting that the standard regularization mechanisms developed for conventional RBHs cannot be directly transferred to string-supported spacetimes. This naturally raises the question of whether a string-supported black hole can be consistently regularized while preserving both the infrared cloud-of-strings behavior and its physical interpretation as an effective string fluid.

It is worth mentioning that the behavior of the CS energy density near the origin, $\rho\sim r^{-2}$, has been associated, for example in reference \cite{Estrada:2026kea}, with so-called integrable singularities (IS) \cite{Lukash:2013ts}. In these geometries, although curvature invariants diverge at the origin, the volume integral of the Ricci scalar remains finite, whereas, for the CS geometry considered here, the corresponding integral of the Kretschmann scalar remains divergent. Furthermore, it has been suggested that, under certain conditions, the tidal forces experienced by freely falling observers may remain finite \cite{Estrada:2023dcj}. See examples of IS in references \cite{Estrada:2024moz,Casadio:2026tmd,Ovalle:2023vvu,Estrada:2026sze}. In recent years, IS have been proposed as an alternative to RBHs models, motivated, among other considerations, by the possible instability of their inner horizons due to the mass-inflation phenomenon \cite{Carballo-Rubio:2022pzu}. However, the stability of RBHs remains a subject of debate. For instance, reference \cite{Bonanno:2022jjp} argue that Hawking evaporation can significantly weaken mass inflation and may even prevent it in certain geometries. More recently, the study in \cite{Bonanno:2025bgc} suggests that the instability associated with mass growth near the Cauchy horizon may be considerably weakened or even avoided under specific conditions. Furthermore, phenomenological models of rotating RBHs exhibiting stability against scalar perturbations have also been constructed in reference \cite{Ghosh:2022gka}. An analysis of RBH stability is beyond the scope of the present work. In this work, we follow the perspective of investigating whether the central singularity of CS geometries can be removed through a regularization constructed under phenomenological assumptions, without claiming to derive such a regularization from a fundamental dynamical theory.

It is therefore of physical interest to investigate whether string-supported black holes can be regularized while preserving both their characteristic infrared behavior and the short-distance features expected from  LQG inspired scenarios such as the Planck star picture. In this direction, it is well known that cloud-of-strings black hole geometries possess a central singularity. A first (and rather naive) approach could consist of applying the usual RBH prescription of replacing the point mass by an effective mass function, $M\rightarrow m(r)$, satisfying $m(r)\sim Cr^3$ near the origin, thereby generating the usual de Sitter core. However, owing to the presence of the cloud-of-strings parameter $\epsilon$, the lapse function behaves as $f(r)\simeq1-\epsilon-C_1r^2$ as $r\rightarrow0$, implying that the dominant contribution to the Kretschmann scalar remains $K\sim4\epsilon^2/r^4$. Hence, the curvature singularity survives even when the mass function satisfies the standard regularity condition. This observation suggests that the ultraviolet structure of string-supported black holes is governed not only by the behavior of the mass function but also by the independent contribution of the string sector, indicating that the conventional de Sitter-core mechanism alone is insufficient to regularize the geometry. Motivated by this observation, rather than introducing another phenomenological regular mass profile, we investigate whether the string sector itself can be consistently screened while retaining its physical interpretation as an effective anisotropic {\it string fluid} satisfying $p_t=W(r)\rho$. The resulting solution preserves the characteristic cloud-of-strings behavior at large distances while yielding a regular geometry at short scales, allowing the same spacetime to consistently interpolate between the infrared regime of string-supported black holes and a nontrivial ultraviolet core. Although the leading near-core geometry remains de Sitter, the string sector introduces subleading corrections to its spatial curvature. This naturally motivates an analysis of the local geometry of the spacelike slices and, in particular, of the corrections to the leading locally $S^3$ geometry associated with de Sitter cores \cite{Bargueno:2020ais,Bargueno:2021fus,Melgarejo:2020mso}.

Since the problem described above cannot be straightforwardly addressed within the standard Einstein field equations, it is natural to resort to a framework capable of systematically incorporating additional gravitational sources while preserving analytical control of the solution. In this regard, the gravitational decoupling (GD) approach has become a powerful nonperturbative framework for constructing compact objects and black hole solutions from known seed geometries in general relativity \cite{Ovalle:2017fgl}. The method consists of solving Einstein's equations for a seed source and subsequently determining the contribution of an additional gravitational sector through a simpler set of quasi-Einstein equations, the complete spacetime being obtained from the combination of both sectors. Recent examples of applications of the gravitational decoupling approach can be found in Refs.~\cite{Estrada:2024lhk,Naseer:2025ghn,Guimaraes:2025jsh,Sallah:2026dhk,Alshammari:2026wnf,Hua:2025qwu}. In the present work, gravitational decoupling is employed not merely as a solution-generating technique, but as a framework for investigating whether the additional string sector responsible for the cloud-of-strings geometry can be consistently regularized while preserving its infrared behavior and admitting a physically meaningful interpretation as an effective anisotropic string fluid. In this way, the method provides a unified description connecting the ultraviolet regularization of the spacetime with the infrared phenomenology characteristic of cloud-of-strings solutions.

Black holes are also thermodynamic systems, characterized by a Hawking temperature and an entropy associated with the horizon area, while the heat capacity provides a diagnostic of their thermal response under specified variations. Since the screening mechanism modifies the spacetime geometry from the vicinity of the horizon down to the core, it is natural to investigate whether these geometric changes induce qualitative modifications in the thermodynamic evolution of the solution. In particular, we analyze the influence of the string-cloud parameter $\epsilon$ on the Hawking temperature, the heat capacity, {temperature turning points}, as well as on the possible formation of black hole remnants. {This analysis also distinguishes the dependence of the extremal mass on $\epsilon$ from the extremal radius, which remains unchanged at fixed regularization scale.} Beyond their thermodynamic properties, the dynamical response of black holes provides an independent probe of the underlying spacetime geometry. Since the screening mechanism introduced in this work modifies both the near-horizon and core regions while preserving the cloud-of-strings asymptotics, it is natural to investigate how these geometric changes are encoded in the quasinormal-mode spectrum. In particular, we analyze whether the string-cloud parameter $\epsilon$ leaves observable imprints on the oscillation frequencies and damping rates of scalar perturbations.

In the present work, we construct a regular black hole supported by an effective string fluid within the framework of gravitational decoupling. The resulting solution provides a regular extension of the cloud-of-strings geometry while preserving its characteristic large-distance behavior, allowing us to investigate how the string sector modifies the horizon structure, thermodynamic evolution, {local spatial geometry of the core}, and quasinormal-mode spectrum. The paper is organized as follows. In Sec.~\ref{sec:Spacetime_Splitting_via_Gravitational_Decoupling}, we construct the regular string-fluid solution through the GD approach. Section~\ref{sec:Event_Horizon_Analysis} is devoted to the analysis of the horizon structure. In Sec.~\ref{sec:Thermodynamics}, we investigate the thermodynamic properties of the solution. Section~\ref{sec:Quasinormal_modes} presents the scalar quasinormal-mode analysis using the sixth-order WKB approximation. Finally, our conclusions are summarized in Sec.~\ref{sec:Discussion_and_conclusions}.

\section{Spacetime Splitting via Gravitational Decoupling}\label{sec:Spacetime_Splitting_via_Gravitational_Decoupling}

We begin by considering a static, spherically symmetric spacetime
in Schwarzschild-like coordinates, whose geometry is characterized
by the radial metric function $f(r)$. The corresponding line
element takes the form

\begin{equation}
ds^{2}
=
-f(r)\,dt^{2}
+\frac{dr^{2}}{f(r)}
+r^{2}\left(d\theta^{2}+\sin^{2}\theta\,d\phi^{2}\right),
\end{equation}
where the metric function is expressed in terms of the effective
mass function $m(r)$ as
\begin{equation}
f(r)=1-\frac{2m(r)}{r}.
\end{equation}

Let us consider that the matter content is described by the energy--momentum tensor of a anisotropic perfect fluid, which in the comoving frame can be written as
\begin{equation}
T^{\mu}{}_{\nu}
=
\mathrm{diag}
\left(
-\rho,\,
p_r,\,
p_t,\,
p_t
\right),
\end{equation}
where $\rho$ is the energy density and $p_r,p_t$ are the radial and tangential pressure, respectively. This leads to the following non-vanishing components of the Einstein field equations,
\begin{align}
8\pi \rho
&=
\frac{1-f(r)-r f'(r)}{r^{2}},
\\
8\pi p_t
&=
\frac{r f''(r)+2f'(r)}{2r},
\end{align}
where it is straightforward to verify that $\rho=-p_r$ and where the prime denotes differentiation with respect to the radial coordinate $r$. The local conservation of the energy--momentum tensor,
\begin{equation}
\nabla_{\mu}T^{\mu}{}_{\nu}=0.
\end{equation}
The above conservation law reduces to
\begin{equation}
\rho'
+
\frac{2}{r}
\left(
\rho+p_t
\right)
=0.
\end{equation}
Therefore, we have three equations, of which only two are independent via the Bianchi identities. Following the gravitational decoupling approach introduced in Ref.\cite{Ovalle:2017fgl}, we apply the Minimal Geometric Deformation (MGD) scheme to the lapse function, which is written as
\begin{equation}
f(r)\rightarrow f_0(r)+\epsilon \cdot \,f_\epsilon(r),
\end{equation}
where $\epsilon$ is the decoupling parameter and $f_\epsilon(r)$ denotes the geometric deformation.
As we shall see below, this naturally induces a decoupling of the gravitational sources of the form
\begin{equation}
T^{\mu}{}_{\nu}
=
\left(T^{\mu}{}_{\nu}\right)_{0}
+
\epsilon\,
\left(T^{\mu}{}_{\nu}\right)_{\epsilon}.
\end{equation}

For simplicity, we shall represent the two gravitational sources in the following form:
\begin{align}
\left(T^{\mu}{}_{\nu}\right)_{0}
&=
\mathrm{diag}
\left(
-\rho_{0},\,
-\rho_{0},\,
(p_{t})_{0},\,
(p_{t})_{0}
\right),
\\
\left(T^{\mu}{}_{\nu}\right)_{\epsilon}
&=
\mathrm{diag}
\left(
-\rho_{\epsilon},\,
-\rho_{\epsilon},\,
(p_{t})_{\epsilon},\,
(p_{t})_{\epsilon}
\right).
\end{align}

It is straightforward to verify that, under this decoupling, the two sources are separately conserved,
\begin{align}
\nabla_{\mu}\left(T^{\mu}{}_{\nu}\right)_0&=0,
\\
\nabla_{\mu}\left(T^{\mu}{}_{\nu}\right)_\epsilon&=0,
\end{align}
which implies that they do not exchange energy or momentum with each other. Their interaction is therefore purely gravitational, through the MGD deformation of the lapse function. Explicitly, the conservation equations for each sector are
\begin{align}
\rho_0'
+
\frac{2}{r}
\left(
\rho_0+(p_t)_0
\right)
&=0,
\\
\rho_\epsilon'
+
\frac{2}{r}
\left(
\rho_\epsilon+(p_t)_\epsilon
\right)
&=0.
\end{align}

As discussed in the Introduction, the standard strategy of replacing the point mass by an effective mass function $M\rightarrow m(r)$ satisfying $m(r)\sim Cr^3$ near the origin, which in regular black hole models leads to a de Sitter core, is not sufficient in the case of geometries supported by a cloud of strings. This is because the presence of the cloud-of-strings parameter $\epsilon$ implies that the lapse function behaves as $f(r)\simeq 1-\epsilon-C_1r^2$ as $r\rightarrow0$. Consequently, the dominant contribution to the Kretschmann scalar satisfies $K\sim 4\epsilon^2/r^4$, revealing that the curvature singularity persists. Motivated by this observation, we shall explore the possibility of modifying the combined gravitational contribution $\frac{2M}{r}+\epsilon$ through a screening factor according to $\frac{2M}{r}+\epsilon \rightarrow \left(\frac{2M}{r}+\epsilon\right)s(r)$, where $s(r)$ denotes a screening function whose form will be determined from the geometric and physical requirements discussed below.
 Thus, a natural choice, following the MGD decomposition is therefore
\begin{align}
f_{0}(r)
&=
1-\frac{2\,G\,M\,s(r)}{r}, \label{f0}
\\
f_{\epsilon}(r)
&=
-s(r), \label{f1}
\end{align}
where $G$ and $M$ denote Newton's gravitational constant and the mass parameter, respectively. In natural units, $[G]=\ell^2$ and $[M]=\ell^{-1}$, where $\ell$ denotes a unit of length. For simplicity, throughout the remainder of this section we set $G=1$. Thus,
\begin{equation} 
f(r)
=f_{0}(r)
+
\epsilon\,f_{\epsilon}(r)
=
1-\frac{2M\,s(r)}{r}
-\epsilon\,s(r) \label{MGD1} =
1-\left (\frac{2M}{r}+\epsilon \right ) s(r)
\end{equation}

This leads, via Einstein equations, to
\begin{align} 
\label{RhoT} \rho_T=\rho_0+\epsilon\, \rho_\epsilon
&=
\frac{M\,s'(r)}{4\pi r^{2}}
+
\epsilon\,
\frac{s(r)+r\,s'(r)}{8\pi r^2},
\\
\label{PtT} (p_t)_T= (p_t)_0+\epsilon\, (p_t)_\epsilon
&=
-\frac{M\,s''(r)}{8\pi r}
-\epsilon\,
\frac{2s'(r)+r\,s''(r)}{16 \pi r}.
\end{align}

Following the idea of the gravitational decoupling algorithm, the previous expressions naturally split into two independent sectors. The zeroth-order $\epsilon$ contributions satisfy the standard Einstein field equations, whereas the first-order $\epsilon$ contributions satisfy a set of quasi-Einstein equations proportional to the decoupling parameter $\epsilon$. In the original formulation of the method, these two systems are referred to as the \emph{Einstein sector} and the \emph{quasi-Einstein sector}, respectively \cite{Ovalle:2017fgl}. Thus:
\begin{align} 
\rho_0
&=
\frac{M\,s'(r)}{4\pi r^{2}},
\\
\epsilon\, \rho_\epsilon&=\epsilon\,
\frac{s(r)+r\,s'(r)}{8\pi r^2}, \\
(p_t)_0
&=
-\frac{M\,s''(r)}{8\pi r},
\\
\epsilon\, (p_t)_\epsilon&= -\epsilon\,
\frac{2s'(r)+r\,s''(r)}{16 \pi r}.
\end{align}

 In the present work, we are particularly interested in investigating whether the quasi Einstein sector, namely the $\mathcal{O}(\epsilon)$ contribution, can effectively describe a string-fluid-like source in some radial domain. To this end, we define effective equations of state for both sectors,
\begin{align}
(p_t)_0
&=
\omega_0(r)\,\rho_0, \label{TangencialCero}
\\
(p_t)_\epsilon
&=
\omega_\epsilon(r)\,\rho_\epsilon. \label{TangencialEpsilon}
\end{align}

We begin by evaluating the equation of state given in
Eq.~\eqref{TangencialEpsilon}, which takes the form
\begin{equation}
r^2 s''(r)
+
2r\left[1+\omega_\epsilon(r)\right]s'(r)
+
2\,\omega_\epsilon(r)s(r)
=
0.
\end{equation}
where $\omega_\epsilon(r)$ characterizes the physical properties
assigned to the quasi-Einstein sector. Thus, rather than claiming
a fundamental dynamical derivation of the screening function
$s(r)$, we determine its physical and analytical properties
under the phenomenological assumptions described below,
requiring that the resulting equations of state described above admit a consistent string-fluid interpretation for the complete geometry and lead to a regular geometry near the origin.  In particular, we shall seek a quasi-Einstein equation of state such that $\omega_\epsilon(r)$ approaches a constant value near the origin. This requirement ensures that the equation of state does not exhibit significant variations in the short-scale regime and allows for a regular power-law behavior of the deformation function. Indeed, if $\omega_\epsilon$ is constant, the above equation reduces to an Euler--Cauchy equation admitting power-law solutions of the form $s(r)\propto r^n$, 
where the exponent $n$ satisfies $n^2+(1+2\omega_\epsilon)n+2\omega_\epsilon=0$. The corresponding roots are
$n=-1,\,\,\, n=-2\omega_\epsilon$, and therefore, $s(r)=C_1\,r^{-1}+C_2\,r^{-2\omega_\epsilon}$. In order to avoid divergences near the origin, we set $C_1=0$ and assume a constant negative value for $\omega_{\epsilon}^{(r\sim0)}$ . In this way,

\begin{equation} \label{SFcercaOrigenOmega}
s(r)\big |_{r \approx 0} \sim C_2\,r^{-2\omega_{\,\,\epsilon}^{(r \sim 0)}}.
\end{equation}

Furthermore, in order for the quasi-Einstein sector to describe a fluid of strings that asymptotically recovers the cloud-of-strings behavior at large scales, we require that $\omega_\epsilon(r)\rightarrow 0$ for $r\rightarrow\infty$. In this limit, the differential equation reduces to
\begin{equation}
r^2s''(r)+2rs'(r)=0,
\end{equation}
whose solution is

\begin{equation}
s(r)=C_4-\frac{C_3}{r} \sim \rightarrow C_4+\mathcal{O}\!\left(\frac{1}{r}\right),
\qquad r\rightarrow\infty.
\end{equation}

That is, if $\omega_\epsilon(r)\to 0$ at large scales, the quasi-Einstein sector asymptotically induces a constant deformation together with a subleading $1/r$ correction. This behavior is consistent with the interpretation of the quasi-Einstein source as a fluid of strings that recovers the cloud-of-strings regime at sufficiently large distances. 
For simplicity, in the following we shall adopt the asymptotic limit $\displaystyle \lim_{r \to \infty}s(r)\rightarrow C_4$. We first note that, in the usual RBH models, the short-scale screening factor behaves as $s(r)\sim r^3$. Since we are interested in configurations for which the undeformed Einstein sector, corresponding to order $\epsilon^0$, describes a regular black hole, a comparison with Eq. \eqref{SFcercaOrigenOmega} naturally leads us to choose

\begin{equation} \label{OmegaConstanteCortaEscala}
\omega_{\epsilon}^{(r\sim 0)}=-\frac{3}{2}.
\end{equation}

As discussed in the Introduction, the Hayward geometry provides an effective description of the Planck-star scenario proposed in Ref.~\cite{DeLorenzo:2014pta}, interpolating between a de Sitter core at short distances and a spacetime whose quantum-gravitational corrections become increasingly suppressed at large scales. Motivated by this behavior, we adopt the following form, inspired by the Hayward screening factor given in Eq. \eqref{EqSF}
\begin{equation}
s(r)=
\frac{r^3}
{L^3\left(LM+\dfrac{r^3}{L^3}\right)}.
\end{equation}
where $L$ is the regularization parameter. In natural units, $[L]=\ell$. This function satisfies $s(r)\sim \frac{r^3}{L^4M}$ for $r\ll L$, which is consistent with the behavior expected from a quasi-Einstein sector characterized by equation \eqref{OmegaConstanteCortaEscala}. While $ \displaystyle \lim_{r \to \infty} s(r)\rightarrow 1$, recovering the constant asymptotic deformation discussed above. Thus, based on the above considerations:

\begin{align}
\rho_0
&=
\rho_H,
\\
\epsilon\,\rho_\epsilon
&=
\frac{\epsilon\,r\left(r^3+4ML^4\right)}
{8\pi\left(r^3+ML^4\right)^2}, \label{DensidadEpsilon}
\\
(p_t)_0
&=
(p_t)_H,
\\
\epsilon\,(p_t)_\epsilon
&=
\frac{3\epsilon\,ML^4\,r\left(r^3-2ML^4\right)}
{8\pi\left(r^3+ML^4\right)^3}, \label{PresionEpsilon}
\end{align}
where $\rho_H$ and $(p_t)_H$ correspond to the generalized energy density and tangential pressure of the Hayward/Planck-star model described in Appendix \ref{ApendicePlanckStar}. This leads to the following equation of state for the quasi-Einstein sector
\begin{equation} \label{EqEstadoOmegaSube}
\omega_\epsilon(r)
=
\frac{(p_t)_\epsilon}{\rho_\epsilon}
=
\frac{
3ML^4\left(r^3-2ML^4\right)
}{
\left(r^3+ML^4\right)\left(r^3+4ML^4\right)
},
\end{equation}
and to the following expressions for the total energy density and pressures
\begin{align}
\rho_T
&=
\rho_H
+
\epsilon\,\frac{\,r\left(r^3+4ML^4\right)}
{8\pi\left(r^3+ML^4\right)^2},
\\
(p_t)_T
&=
(p_t)_H
+
\epsilon\,\frac{3\,ML^4\,r\left(r^3-2ML^4\right)}
{8\pi\left(r^3+ML^4\right)^3}.
\end{align}
It is worth noting that, using Eqs.~\eqref{RhoT}, \eqref{PtT}, \eqref{TangencialCero}, and \eqref{TangencialEpsilon}, it is also possible to define a global equation of state for the Einstein and quasi-Einstein sectors,
\begin{equation} 
(p_t)_T=W(r) \,\rho_T,
\end{equation}
which can be written explicitly as
\begin{equation} \label{EqEstadoTotal}
W(r)=
\frac{(p_t)_0+\epsilon\cdot \omega_\epsilon(r)\rho_\epsilon}
{\rho_0+\epsilon \cdot \rho_\epsilon}.
\end{equation}
then
\begin{equation} \label{EqEstadoTotalReemplazada}
W(r)=
\frac{
(p_t)_H
+
\epsilon\,\dfrac{3\,ML^4\,r\left(r^3-2ML^4\right)}
{8\pi\left(r^3+ML^4\right)^3}
}{
\rho_H
+
\epsilon\,\dfrac{\,r\left(r^3+4ML^4\right)}
{8\pi\left(r^3+ML^4\right)^2}
}.
\end{equation}
The physical implications of the above equations of state in both the short- and long-distance regimes will be discussed below.

\subsection{ Long-distance regime:}

Introducing the dimensionless variable $x=\frac{M}{r}$.
As noted in Appendix~I, in this regime, for $x\ll 1$ (equivalently, $r\gg M$),  both the energy density and the tangential pressure of the Hayward seed source, Eqs.~\eqref{DensidadCeroLargaEscala} and \eqref{TangencialCeroLargaEscala}, are of order $x^6$. This indicates that the quantum corrections to the Hayward seed density become almost negligible at large scales. Nevertheless, they may still remain marginally testable over a few characteristic length scales before the $r^{-1}$ contribution in the metric function \eqref{LQG} becomes dominant. This behavior is consistent with the expectation that effects associated with Loop Quantum Gravity are more significant at short distances, as discussed in the Introduction.

We are interested in analyzing the quasi Einstein sector sector in this regime from eq \eqref{DensidadEpsilon}

\begin{equation} \label{DensidadEpsilonLargaEscala}
\epsilon \, \rho_\epsilon =
\frac{\epsilon}{8\pi M^2}\,
\frac{x^2+\dfrac{4L^4}{M^2}x^5}
{\left(1+\dfrac{L^4}{M^2}x^3\right)^2}
\approx
\frac{\epsilon}{8\pi M^2}\,x^2
+\frac{\epsilon L^4}{4\pi M^4}\,x^5
+\mathcal{O}(x^8)
\approx \frac{\epsilon}{8\pi r^2},
\end{equation}
which, as we discuss below, behaves as the energy density associated with a cloud of strings. The equation-of-state parameter \eqref{EqEstadoOmegaSube}, $\omega_\epsilon$, of the quasi-Einstein sector behaves in this regime as
\begin{align}
\omega_\epsilon(x) &=\frac{
3\dfrac{L^4}{M^2}x^3-
6\dfrac{L^8}{M^4}x^6
}{
1
+
5\dfrac{L^4}{M^2}x^3
+
4\dfrac{L^8}{M^4}x^6
}
\approx
3\frac{L^4}{M^2}x^3
+
\mathcal{O}(x^6)=
3\frac{L^4M}{r^3} 
+ \mathcal{O}\!\left(\frac{M^6}{r^6}\right) \approx \mathcal{O}(x^3)\label{OmegaLargaEscalaEpsilon}.
\end{align}

Replacing Eqs.~\eqref{DensidadEpsilonLargaEscala} and \eqref{OmegaLargaEscalaEpsilon} into Eq. of state \eqref{TangencialEpsilon}, we obtain

\begin{equation}
    (p_t)_\epsilon \approx \mathcal{O}\left(x^5\right).
\end{equation}

Furthermore, we note that in this regime, by using Eqs. \eqref{SFLargaEscala} and \eqref{MGD1}, the metric tensor takes the following form:
\begin{equation}
f(r)\approx
1-\epsilon-\frac{2M}{r}
+\epsilon\,\frac{L^4M}{r^3}
+\frac{2L^4M^2}{r^4}
+\mathcal{O}(r^{-6}).
\end{equation}

Thus, at large scales, we find that the equation of state of the quasi Einstein sector is such that its energy density behaves precisely as that of a string cloud, as given by Eq.~\eqref{DensidadEpsilon}, while both the equation-of-state parameter $\omega_\epsilon(r)$ and the tangential pressure of this sector, $p_\epsilon$, vanish. In this way, at large scales the deformative sector exhibits a behavior mimicking a string cloud. Therefore, we can state that the energy density, tangential pressure, and the function $\omega_\epsilon(r)$ can effectively represent a string fluid. We also note that, in this regime, the behavior of the quantities discussed above reveals an LQG-like correction of the form $\sim \epsilon\, r^{-3}$. This contribution is, in principle, more easily testable at large distances than the Hayward correction mentioned in the Introduction. Nevertheless, as previously argued, corrections of quantum-geometric origin are expected to become more pronounced at short scales due to their LQG nature, as we discuss below. By evaluating the global equation of state, Eq.~\eqref{EqEstadoTotal}, we find that
\begin{equation}
W(r)
\approx
\frac{
\mathcal{O}(x^6)
+
\left(3\dfrac{L^4}{M^2}x^3\right)
\left(\dfrac{\epsilon}{8\pi M^2}x^2\right)
}{
\mathcal{O}(x^6)
+
\dfrac{\epsilon}{8\pi M^2}x^2
}
\approx
3\frac{L^4}{M^2}x^3
+\mathcal{O}(x^4)
\sim \mathcal{O}(x^3),
\end{equation}

Therefore, at large distances the effective equation of state of the total source approaches that of a cloud of strings, characterized by a nonvanishing energy density and a vanishing tangential pressure.

\subsection{Short--scales regime:} 

Introducing the dimensionless variable $x=\frac{r}{L}$.
We note that in the short-scale regime, $x\ll 1$ (equivalently, $r\ll L$), the seed equation of state satisfies $\omega_0 \approx -1$, as follows from Eq.~\eqref{OmegaCeroCortaH}. This behavior resembles that of a de Sitter core. Indeed, the energy density $\rho_0$ approaches a constant value in this limit, as shown in Eq.~\eqref{DensidadCeroCorta}, while, consistently with the above discussion, the tangential pressure satisfies $(p_t)_0 \approx -\rho_0$, as given in Eq.~\eqref{DeSitterCorta}. We are interested in analyzing the quasi Einstein sector in this regime, for $x\ll 1$ (equivalently, $r\ll L$). From Eq.~\eqref{DensidadEpsilon}, one finds
\begin{align}
\epsilon\,\rho_\epsilon
&\approx
\frac{\epsilon}{8\pi L^2}
\frac{x(x^3+4ML)}
{(x^3+ML)^2}
\approx
\frac{\epsilon}{2\pi ML^3}x
+\mathcal{O}(x^4)
=
\frac{\epsilon r}{2\pi ML^4}
+\mathcal{O}(r^4).
\label{DensidadEpsilonCortaEscala}
\end{align}

The equation-of-state parameter of the quasi-Einstein sector behaves as
\begin{align}
\omega_\epsilon(x)
&=
-\frac{
6LM\left(LM-\dfrac{x^3}{2}\right)
}{
4L^2M^2+5LMx^3+x^6
}  \approx -\frac{3}{2} +
\mathcal{O}(x^3).
\label{OmegaCortaEscalaEpsilon}
\end{align}

Replacing Eqs.~\eqref{DensidadEpsilonCortaEscala} and
\eqref{OmegaCortaEscalaEpsilon} into Eq.~\eqref{TangencialEpsilon}, we obtain
\begin{equation}
(p_t)_\epsilon
\approx
-\frac{3\,r}{4\pi ML^4}
+
\mathcal{O}\!\left(\frac{r^4}{L^8}\right).
\end{equation}

It is also instructive to examine the tangential null
energy condition (NEC), $\rho_T+(p_t)_T\geq 0$,
in the near-center regime. For the Hayward case
($\epsilon=0$), the leading contributions to the
energy density and tangential pressure cancel,
as follows from Eqs.~\eqref{DensidadCeroCorta}
and \eqref{TangencialCeroCorta}. However,
retaining the first nonvanishing contribution,
we obtain $\rho_H+(p_t)_H
\approx
\frac{9r^3}{4\pi ML^8}>0.$ . Thus, the tangential NEC is satisfied in the
Hayward geometry, although its value becomes
arbitrarily small near the origin, vanishing
at order $\mathcal{O}(r^3)$. 
For $\epsilon\neq0$, the string sector modifies
this behavior. Combining the contributions
from both gravitational sectors, we obtain 
\begin{equation}
\rho_T+(p_t)_T
\approx
-\frac{\epsilon r}{4\pi ML^4}
+\frac{9r^3}{4\pi ML^8}
+\mathcal{O}(r^4).
\end{equation}
For $\epsilon>0$, the negative linear term
dominates sufficiently close to the origin,
leading to a violation of the tangential NEC
in a neighborhood of the center, in contrast
with the Hayward case. Therefore, although
the string contribution preserves the leading
de Sitter behavior, it introduces a linear
correction that leads to a near-center tangential
NEC violation absent in the Hayward geometry.

On the other hand, the global equation of state becomes
\begin{equation}
W(r)
\approx
\frac{
-\dfrac{3}{4\pi L^4}
-\dfrac{3\epsilon}{4\pi ML^3}\,x
}{
\dfrac{3}{4\pi L^4}
+\dfrac{\epsilon}{2\pi ML^3}\,x
}
=
-\frac{3\left(1+\dfrac{\epsilon}{ML}x\right)}
{3+\dfrac{2\epsilon}{ML}x}.
\end{equation}

Therefore,
\begin{equation} \label{EqEstadoGlobalCorta}
W(r)
\approx
-1-\frac{\epsilon L}{3M}x
+\mathcal{O}(x^2)
=
-1-\frac{\epsilon r}{3M}
+\mathcal{O}(r^2),
\qquad
x=\frac{r}{L}\ll1.
\end{equation}

In this way, the presence of the cloud-of-strings sector,
controlled by the parameter $\epsilon$, modifies the
effective equation of state of the total source at order $r$.
As a consequence, the spacetime no longer exhibits an
exact de Sitter core in the central region for
$\epsilon\neq 0$. While the undeformed Hayward sector
satisfies $(p_t)_0\simeq-\rho_0$ near the origin,
the additional cloud-of-strings contribution shifts
the effective equation-of-state parameter away from $-1$,
yielding instead the behavior given by
Eq.~\eqref{EqEstadoGlobalCorta}.  Furthermore, we note that in this regime, by using
Eqs.~\eqref{dSCore} and \eqref{MGD1}, the metric function takes the
following form:
\begin{equation}
f(r)
\approx
1
-
\frac{2r^2}{L^4} -\epsilon\,
\frac{r^3}{L^4M}
+
\mathcal{O}(r^5).
\end{equation}

We can define an effective de Sitter radius through $L_{\rm eff}^{2}=\frac{L^{4}}{2}$. Equivalently, the effective cosmological constant is given by $\Lambda_{\rm eff}=\frac{3}{L_{\rm eff}^{2}}=\frac{6}{L^{4}}$. Therefore, the metric function becomes

\begin{equation}
f(r)
\approx
1-\frac{r^2}{L_{\rm eff}^{2}}
-\frac{\epsilon}{L^{4}M}r^{3}
+\mathcal{O}(r^{5}).
\end{equation}

Thus, at short scales, the quasi-Einstein sector introduces non-trivial modifications to the original Hayward core. In particular, the total energy density is no longer approximately constant, but acquires a leading linear correction in the radial coordinate. As a consequence, the central region no longer behaves as an effective positive cosmological constant, signaling a departure from the standard de Sitter structure of the seed solution.

This change is also encoded in the effective equation of state of the quasi-Einstein sector. Hence, the gravitational decoupling mechanism modifies not only the geometry but also the effective matter content supporting the regular interior. Furthermore, the metric develops an additional correction of the form $\sim \epsilon \,r^3$, which becomes increasingly relevant as the Planckian region is approached. This contribution is consistent with the disappearance of the exact de Sitter core and with the emergence of an additional repulsive potential generated by the deformative sector. Therefore, in agreement with its LQG-inspired origin, the effects associated with the gravitational deformation become progressively more important at short distances. Since the resulting core is no longer exactly de Sitter, it is natural to examine whether the spacetime nevertheless remains free of curvature singularities. To address this point, we now study the behavior of the Kretschmann scalar in the short-scale regime. The Ricci scalar is found to be

\begin{equation} \label{RicciCortaEscala}
R(r)
=
\frac{12}{L_{\rm eff}^{2}}
+
\frac{20\epsilon}{L^{4}M}\,r
+
\mathcal{O}(r^{2}) \approx R_{\rm dS} + \epsilon\cdot \frac{20}{L^{4}M}\,r\,.
\end{equation}

Similarly, the Kretschmann scalar behaves as

\begin{equation}
K(r)
=
\frac{24}{L_{\rm eff}^{4}}
+
\frac{80\epsilon}{L_{\rm eff}^{2}L^{4}M}\,r
+
\mathcal{O}(r^{2}) \approx K_{\rm dS}
+
\epsilon \cdot \frac{80}{L_{\rm eff}^{2}L^{4}M}\,r \, .
\end{equation}

Therefore, both invariants remain finite at $r=0$, confirming that the core is regular and free of curvature singularities. The quasi-Einstein sector induces a linear correction in $r$, proportional to the decoupling parameter $\epsilon$. 

\subsection{Near-core spatial geometry and its deformation}

The topology of regular black holes has been studied
in the literature. Borde's theorem established that
spherically symmetric regular black holes satisfying
suitable global conditions are necessarily associated
with a change in topology \cite{Borde:1996df}.
Motivated by this result, Ref.~\cite{Bargueno:2020ais}
showed that regular black holes with de Sitter
(anti-de Sitter) cores possess a local $S^3$ ($H^3$)
spatial geometry. Subsequently,
Ref.~\cite{Melgarejo:2020mso} provided a geometric
interpretation, through Carter--Penrose diagrams,
of the topological transition of spacelike slices
from $\mathbb{R}\times S^2$ outside the regular
black-hole core to $S^3$ in its inner region,
under suitable spacetime identifications,
with a locally $S^3$ geometry near the center. Reference  \cite{Bargueno:2021fus} extended this analysis
to other classes of regular cores. In the present regular fluid-of-strings solution,
the leading near-core geometry remains de Sitter.
However, the screening induced through the MGD
procedure introduces subleading corrections
proportional to the string-cloud parameter
$\epsilon$, modifying the local spatial curvature
away from the center. It is therefore natural to
investigate how this contribution deforms the
near-core spatial geometry while preserving its
leading $S^3$ structure, without assuming that
such a deformation implies a change in global
topology. To this end, we follow the geometric
procedure developed in
Refs.~\cite{Bargueno:2020ais,Bargueno:2021fus,Melgarejo:2020mso}.
For the analysis presented in this subsection,
we follow the procedure developed in
Ref.~\cite{Melgarejo:2020mso}.
We introduce a coordinate system on the
spherically symmetric spacelike hypersurface
such that the line element can be written as
\begin{equation}
\label{Lamina}
dl^2=
\frac{r^2}{\lambda(r,\epsilon)}\,dr^2
+
r^2d\Omega^2.
\end{equation}
where, as shown in Ref.~\cite{Melgarejo:2020mso}, the spatial geometry near the radial origin can be characterized through the three-dimensional Ricci scalar,
\begin{equation}
R^{(3D)}(r,\epsilon)
=
\frac{2\left[
r\left(r-\partial_r\lambda(r,\epsilon)\right)
+\lambda(r,\epsilon)
\right]}{r^4}.
\end{equation}
thus, imposing regularity at the origin, the corresponding solution can be expressed as
\begin{equation}
\lambda(r,\epsilon)
=
r\int_0^r
\left[
1-\frac{y^2}{2}R^{(3)}(y,\epsilon)
\right]dy.
\end{equation}

For our solution, the spatial Ricci scalar naturally separates into a seed contribution and a correction proportional to the string-cloud parameter,
\begin{equation}
R^{(3D)}(r,\epsilon)
=
R^{(3D)}_{\mathrm{seed}}(r)
+
\epsilon R^{(3D)}_{\epsilon}(r),
\end{equation}
where, near the origin,
\begin{equation}
R^{(3D)}_{\mathrm{seed}}(r)
=
\frac{12}{L^4}
+\mathcal{O}(r^3)\,\,\,,
\,\,
R^{(3D)}_{\epsilon}(r)
=
\frac{8r}{ML^4}
+\mathcal{O}(r^4).
\end{equation}
thus, consequently, the regular near-core expansion of $\lambda(r,\epsilon)$ becomes
\begin{equation}
\lambda(r,\epsilon)
=
r^2
-
\frac{R^{(3D)}_{\mathrm{seed}}(0)}{6}r^4
-
\frac{\epsilon}{8}
\left(R^{(3D)}_{\epsilon}\right)'(0)r^5
+
\mathcal{O}(r^7).
\end{equation}

Since the leading curvature term in the near-core
expansion of $\lambda(r,\epsilon)$ is determined
by $R^{(3)}_{\mathrm{seed}}(0)$,
whereas the string-cloud contribution first
appears at higher order, for simplicity,
we adopt the leading-order approximation
to the spatial geometry near the origin
and introduce the following coordinate
transformation,
\begin{equation}
1-
\frac{r^2R^{(3D)}_{\mathrm{seed}}(0)}{6}
\equiv
\cos^2\xi.
\end{equation}
thus, since $R^{(3D)}_{\mathrm{seed}}(0)=12/L^4$, one finds $\sin^2\xi=2r^2/L^4$ and $r=\frac{L^2}{\sqrt{2}}\sin\xi$. Then, using the complete near-core expansion of $\lambda(r,\epsilon)$, we obtain
\begin{equation}
\frac{\lambda(r,\epsilon)}{r^2}
=
\cos^2\xi
-
\epsilon\,\frac{ L^2}{2\sqrt{2}M}\sin^3\xi
+
\mathcal{O}(\sin^5\xi).
\end{equation}
therefore, the spatial metric \eqref{Lamina} near the origin becomes
\begin{equation}
\label{TopologiaTotal}
\begin{aligned}
dl^2={}&
\frac{L^4}{2}
\frac{
\cos^2\xi
}{
\cos^2\xi
-
\epsilon\,\dfrac{ L^2}{2\sqrt{2}M}\sin^3\xi
+
\mathcal{O}(\sin^5\xi)
}
\,d\xi^2
\\
&+
\frac{L^4}{2}
\sin^2\xi\,d\Omega^2.
\end{aligned}
\end{equation}

We first consider the undeformed case $\epsilon=0$,
for which, at leading order near the origin,
\begin{equation}
dl^2
\simeq
\frac{L^4}{2}
\left(
d\xi^2
+
\sin^2\xi\,d\Omega^2
\right).
\end{equation}
This corresponds to a locally $S^3$ spatial geometry,
consistent with Ref.~\cite{Melgarejo:2020mso},
where a global transition from
$\mathbb{R}\times S^2$ to $S^3$ was established
for the spacetime extensions considered therein.

For $\epsilon\neq0$, Eq.~\eqref{TopologiaTotal}
shows that the string-cloud contribution
introduces a subleading deformation of the
near-core spatial geometry, proportional to
$\epsilon\sin^3\xi$.
Although this correction does not modify
the leading local $S^3$ spatial geometry,
it introduces a nontrivial subleading
correction to the three-dimensional Ricci scalar,
while preserving the leading de Sitter
behavior near the origin.
Nevertheless, this smooth geometric
deformation does not, by itself,
establish a change in the global topology
of the spacelike hypersurfaces, which
would require an analysis of the global
spacetime extension beyond the scope
of this work.

\section{Event Horizon Analysis}\label{sec:Event_Horizon_Analysis}

We now analyze the horizon structure of the solution. The horizons are determined by the zeros of the lapse function $f(r)$, corresponding to the hypersurfaces where the Killing vector $\partial_t$ becomes null. Depending on the parameters, these zeros describe an inner horizon $r_i$ and an event horizon $r_h$, with $r_h>r_i$. The horizon equation is obtained from the condition $f(r)=0$,

\begin{equation}
(1-\epsilon)r^{3}-2GMr^{2}+L^{4}M=0.
\end{equation}

We solve the previous equation by following a mathematical procedure analogous to that of Ref \cite{Halilsoy:2013iza}. For the remainder of this section, we set $G=1$ for simplicity. Introducing the effective mass parameter $m\equiv M/(1-\epsilon)$, this equation can be written as

\begin{equation}
r^{3}-2mr^{2}+L^{4}m=0.
\label{eq:cubic}
\end{equation}

Although Eq.~\eqref{eq:cubic} resembles the algebraic form of the corresponding Hayward horizon equation, the presence of the string-cloud parameter $\epsilon$ gives rise to a physically distinct spacetime. As we will see below, it changes the short- and long-distance behavior of the geometry, the matter content, the global equation of state, and the thermodynamic properties of the solution. To analyze the roots of the cubic equation \eqref{eq:cubic}, we introduce the dimensionless variable $\bar{\rho}$ via $r = m\,\bar{\rho}$, such that the horizon equation takes the compact form
\begin{equation}
\bar{\rho}^{3}-2\bar{\rho}^{2}+2\lambda^{2}=0, \label{eq:dimless}
\end{equation}
where we define the parameter $\lambda^{2} \equiv L^{4}/2m^{2}$. The nature of the roots of this cubic is determined by its discriminant. For the cubic equation under consideration, the discriminant can be computed explicitly and is given by
\begin{equation}
\mathcal{D}=4\lambda^{2}\left(16-27\lambda^{2}\right). \label{eq:discriminant}
\end{equation}

The sign of the discriminant $\mathcal{D}$ completely determines the horizon structure. For $\mathcal{D}>0$ ($\lambda^{2}<16/27$), the cubic has two positive roots corresponding to the inner and event horizons. For $\mathcal{D}=0$ ($\lambda^{2}=16/27$), these horizons merge into a degenerate extremal horizon. Finally, for $\mathcal{D}<0$ ($\lambda^{2}>16/27$), no positive root exists and the solution describes a regular horizonless compact object. Defining $\lambda_{\rm cri}^{2}=\frac{16}{27}$, the above remarks can be summarized as follows:\begin{equation}
\left\{
\begin{aligned}
\lambda^{2}<\lambda_{\rm cri}^{2}
&\quad\Longrightarrow\quad
\text{inner horizon }(r_i)\ \text{and event horizon }(r_h),
\\[2mm]
\lambda^{2}=\lambda_{\rm cri}^{2}
&\quad\Longrightarrow\quad
\text{extremal black hole }(r_i=r_h),
\\[2mm]
\lambda^{2}>\lambda_{\rm cri}^{2}
&\quad\Longrightarrow\quad
\text{regular horizonless compact object.}
\end{aligned}
\right.
\end{equation}

\begin{figure}[ht]
\centering
\includegraphics[width=0.7\textwidth]{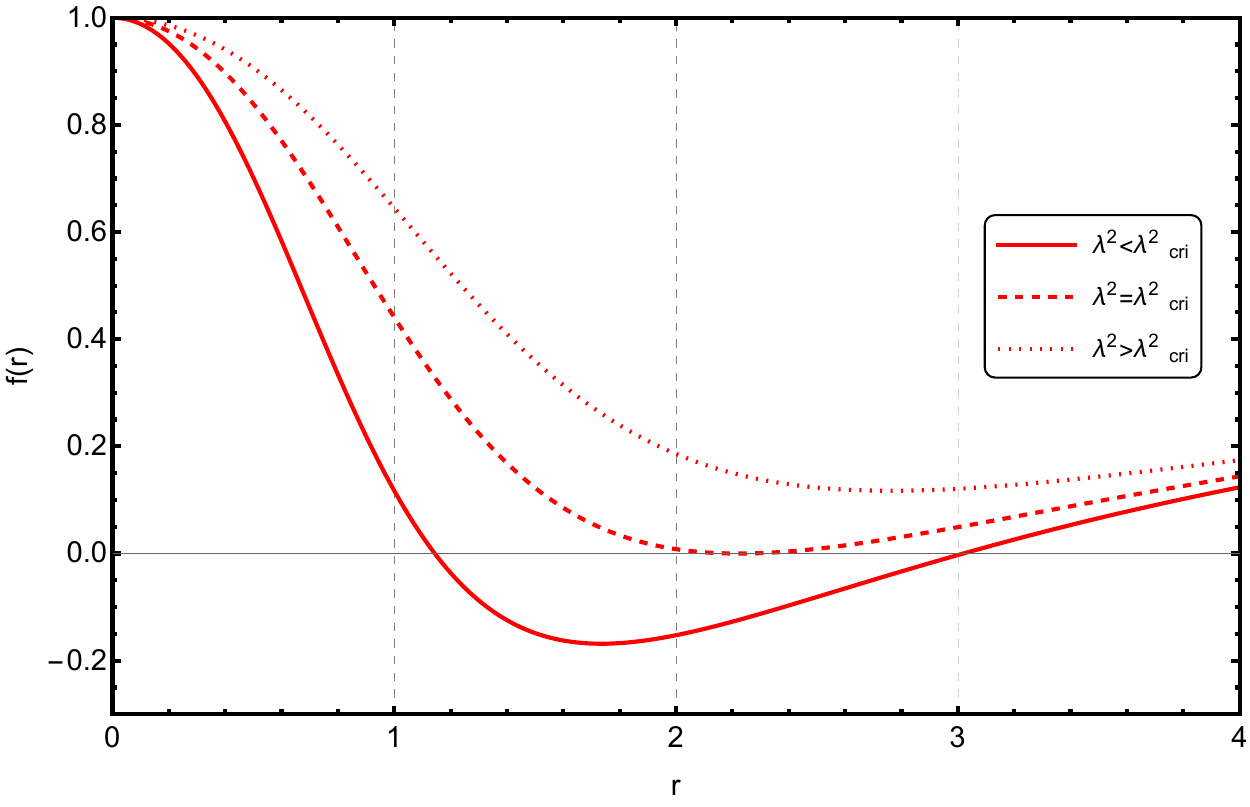}
\includegraphics[width=0.7\textwidth]{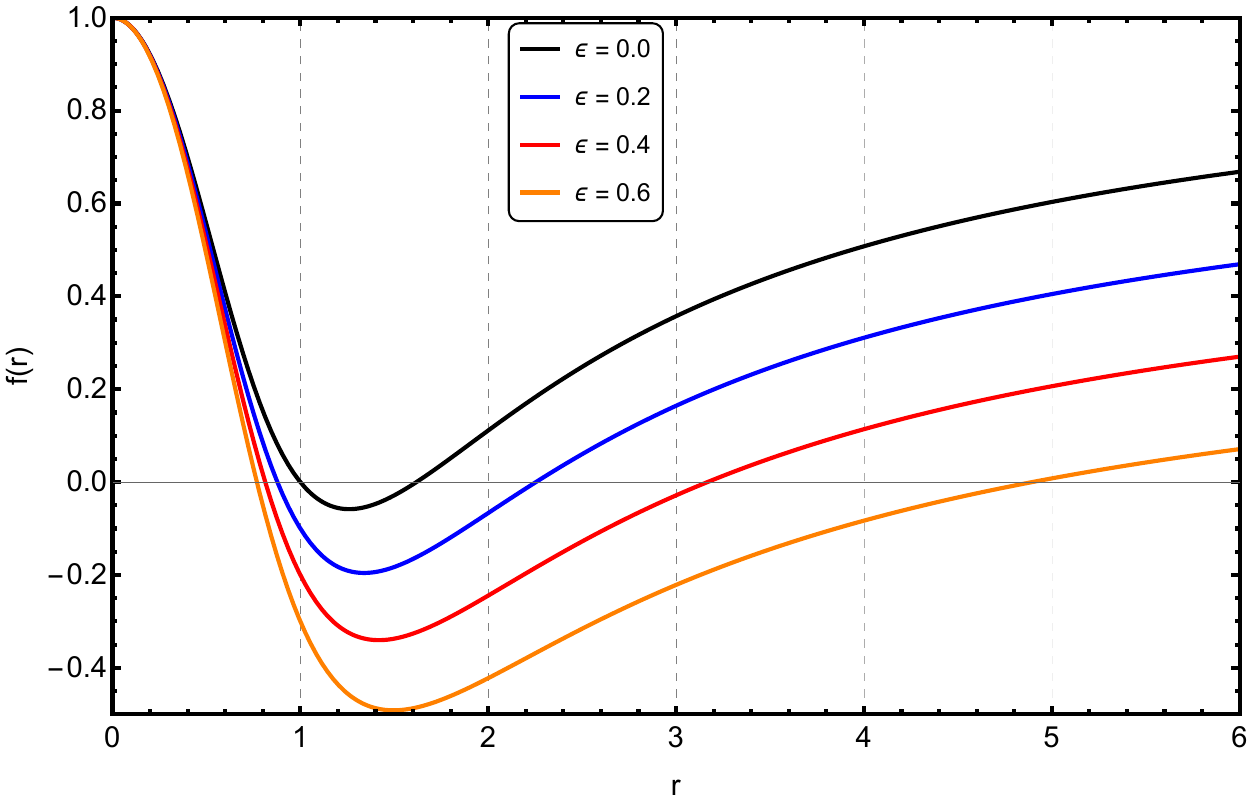}
\caption{Lapse function $f(r)$. \textbf{First panel:} $L=1$ and $\epsilon=0.4$, illustrating the three horizon configurations corresponding to $\lambda>\lambda_{\rm cri}$, $\lambda=\lambda_{\rm cri}$, and $\lambda<\lambda_{\rm cri}$. \textbf{Second panel:} Lapse function $f(r)$ for selected values of the string-cloud parameter $\epsilon$, with $M=L=1$. As $\epsilon$ increases, the event horizon shifts towards larger radii, while the asymptotic value of the lapse function approaches $1-\epsilon$.}
\label{fig:f_r}
\end{figure}

The behavior described above is illustrated in the first panel of Fig.~\ref{fig:f_r}, which shows the three possible horizon configurations corresponding to $\lambda>\lambda_{\rm cri}$, $\lambda=\lambda_{\rm cri}$, and $\lambda<\lambda_{\rm cri}$. The second panel displays the effect of the string-cloud parameter $\epsilon$. The zeros of $f(r)$ identify the inner and event horizons. As $\epsilon$ increases, the event horizon shifts towards larger radii, while the asymptotic value of the lapse function approaches $f(r)\rightarrow1-\epsilon$.

The largest positive root of Eq.~\eqref{eq:dimless}, corresponding to the event horizon in the non-extremal case, is obtained from Cardano's formula

\begin{equation}
\bar{\rho}_{h}=
\frac{1}{3}
\left(
\sqrt[3]{\Delta}
+\frac{4}{\sqrt[3]{\Delta}}
+2
\right),
\label{eq:rho_h}
\end{equation}
where

\begin{equation}
\Delta=
8-27\lambda^{2}
+
\sqrt{27\lambda^{2}
\left(27\lambda^{2}-16\right)}.
\label{eq:Delta}
\end{equation}

Using $r=m\bar{\rho}$, with $m=M/(1-\epsilon)$,
the event horizon radius becomes

\begin{equation}
r_{h}=
\frac{M}{3(1-\epsilon)}
\left(
\sqrt[3]{\Delta}
+\frac{4}{\sqrt[3]{\Delta}}
+2
\right).
\label{eq:r_h}
\end{equation}

{For $0<\lambda^2\leq16/27$,
the principal complex cube root of $\Delta$
in Eq.~\eqref{eq:rho_h} yields a real
outer event horizon.}
Substituting
$\lambda^{2}=\frac{L^{4}}{2m^{2}}
=\frac{(1-\epsilon)^{2}L^{4}}{2M^{2}}$
into Eq.~\eqref{eq:Delta}, we obtain

\begin{equation}
\Delta=
8
-
\frac{27(1-\epsilon)^{2}L^{4}}{2M^{2}}
+
\sqrt{
\frac{27(1-\epsilon)^{2}L^{4}}{2M^{2}}
\left(
\frac{27(1-\epsilon)^{2}L^{4}}{2M^{2}}
-16
\right)
}.
\label{eq:Delta_full}
\end{equation}

Therefore, the event horizon radius is completely determined by the parameters $M$, $L$, and $\epsilon$.

The extremal configuration is obtained from $\lambda^{2}=16/27$, yielding $
L_{E}^{4}=\frac{32M^{2}}{27(1-\epsilon)^{2}}$. Thus, the corresponding double root is $\bar{\rho}_E=4/3$, giving

\begin{equation}
r_{h,E}
=
\frac{4M}{3(1-\epsilon)}
\;\Longrightarrow\;
2r_{h,E}^{2}=3L^{4}.
\label{eq:r_E}
\end{equation}

Hence, for fixed $M$, the string-cloud parameter affects both the extremal regularisation scale and the extremal horizon radius. As $\epsilon\rightarrow1^{-}$, both quantities increase monotonically according to $L_{E}^{4}\propto(1-\epsilon)^{-2}$ and $r_{h,E}\propto(1-\epsilon)^{-1}$. {For fixed $L$, however, the extremal radius $r_{h,E}=L^2\sqrt{3/2}$ is independent of $\epsilon$.}

\section{Thermodynamics}\label{sec:Thermodynamics}

We now investigate the thermodynamic properties
of our RBH geometry. We first clarify the normalization of the
Killing vector and its implications for the
thermodynamic quantities.

\subsection{{Killing Vector Normalization}}

Since $s(r)\to1$ as $r\to\infty$, the metric
function \eqref{MGD1} approaches $f(r)\to1-\epsilon$.
Therefore, the geometry is not asymptotically flat.
Consequently, the usual Killing vector
$\xi=\partial_t$ is not unit-normalized
at infinity. For $0\leq\epsilon<1$,
the redefinition of the temporal coordinate
$\widehat{t}=\sqrt{1-\epsilon}\,t$
leads to the normalized timelike Killing vector
$\widehat{\xi}=\xi/\sqrt{1-\epsilon}$,
such that $\displaystyle \lim_{r\to\infty}
g_{\mu\nu}\widehat{\xi}^{\mu}
\widehat{\xi}^{\nu}=-1$. The surface gravity associated with the normalized
Killing vector satisfies $\widehat{\xi}^{\nu}\nabla_\nu
\widehat{\xi}^{\mu}
=\widehat{\kappa}\widehat{\xi}^{\mu}$. Using $\widehat{\xi}^{\mu}
=\xi^{\mu}/\sqrt{1-\epsilon}$
and $\xi^\nu\nabla_\nu\xi^\mu=\kappa\xi^\mu$, we obtain
\begin{equation}
\widehat{\kappa}
=\frac{\kappa}{\sqrt{1-\epsilon}} \Rightarrow \widehat{T}=\frac{\widehat{\kappa}}{2\pi}
=\frac{T}{\sqrt{1-\epsilon}}
\end{equation}
where $\kappa$ and $T$ denote the surface gravity
and Hawking temperature associated with the original
Killing generator $\xi=\partial_t$ and the original
coordinate time $t$.
For simplicity, throughout the following analysis,
we retain this convention and report the temperature
$T$, which differs from $\widehat{T}$ by the constant
factor $\sqrt{1-\epsilon}$. The implications of this normalization for the
quasinormal frequencies are discussed separately
in the corresponding section.

\subsection{{Temperature and Thermal Response}} \label{DiscusionNormalizacion}

We now investigate the thermodynamic properties of our fluid of strings black hole geometry. For simplicity, we set $G=1$. In accordance with the preceding discussion, the Hawking temperature is obtained from the surface gravity at the event horizon through the standard relation
\begin{equation}
    T = \frac{\kappa}{2\pi}
      = \left.\frac{f'(r)}{4\pi}\right|_{r=r_h}.
\end{equation}

Evaluating the derivative of the lapse function and substituting the mass parameter $M(r_h)$ determined by the horizon condition $f(r_h,M)=0$, we obtain
\begin{equation}
    T =
    \frac{(1-\epsilon)\left(2r_h^{2}-3L^{4}\right)}
         {4\pi r_h\left(2r_h^{2}-\epsilon L^{4}\right)}.
    \label{eq:temp}
\end{equation}

We note that the Hawking temperature vanishes at the
horizon radius $r_{h,E}^{2}=(3/2)L^{4}$, given by
Eq.~\eqref{eq:r_E}, which is independent of the string
parameter $\epsilon$ for fixed $L$.
Since the Hawking temperature is proportional to the
derivative of the metric function at the horizon,
$T_H\propto f'(r_h)$, its vanishing at $r_{h,E}$
is associated with the degeneracy of the horizon,
characterized by $f(r_{h,E})=f'(r_{h,E})=0$.
Therefore, this configuration corresponds to an
extremal black hole, with an event horizon radius of $r_{h,E}$ and vanishing Hawking temperature. 
Note that the standard thermal Hawking emission vanishes
in the extremal limit. Nevertheless, extremal black holes
may still possess nonzero entropy, associated with the
number of quantum microstates of the system
\cite{Estrada:2025lgo}.
As discussed below, this extremal solution
may be interpreted as a possible black-hole remnant,
representing a limiting configuration toward which
an evaporating black hole could evolve \cite{Adler:2001vs}.
However, establishing whether evaporation actually
terminates at this configuration requires a dynamical
analysis beyond the scope of the present work.

Furthermore, for event horizon radii larger than the
extremal value, $r_h>r_{h,E}$, we have $2r_h^2>2r_{h,E}^2=3L^4>\epsilon L^4$, since $0\leq\epsilon<1$. Therefore, it follows directly
from the expression for the Hawking temperature that $T>0$ throughout the non-extremal black hole branch. Furthermore, in the limit $\epsilon \to 0$, this reduces to the standard Hayward temperature.  

It is well known that the first law of thermodynamics can be modified for regular black holes (RBHs) due to the presence of matter fields in the energy-momentum tensor \cite{Ma:2014qma}. To investigate a structure analogous to the first law, we consider the horizon conditions $f(r_h,M)=0$ and $\delta f(r_h,M)=0$, keeping $L$ and $\epsilon$ fixed. These conditions constrain variations in the parameter space and lead to $-\frac{r_h}{2}
\left.\frac{\partial f}{\partial M}\right|_{r=r_h}dM
=
\left.\frac{1}{4\pi}
\frac{\partial f}{\partial r}\right|_{r=r_h}
(2\pi r_h\,dr_h)$, which takes the following form:

\begin{equation}
    du=\mathcal{F}(r_h)\,dM= T\,dS,
    \qquad
    \mathcal{F}(r_h)=
    \frac{\left(2r_h^2-L^4\right)^2}
         {2r_h^2\left(2r_h^2-\epsilon L^4\right)}.
    \label{eq:reduced_first_law}
\end{equation}
where $\mathcal{F}(r_h)$ follows directly from the
horizon equation and acts as a multiplicative factor modifying the mass variation in the thermodynamic relation. Accordingly, the combination
$\mathcal{F}(r_h)\,dM$, following Ref.~\cite{Estrada:2020tbz,Estrada:2023cyx}, may be interpreted as an
effective and local energy variation $du$, such that
$du=T\,dS$.

The local thermodynamic stability of the black hole is determined by the sign of the heat capacity at fixed $L$ and $\epsilon$. The sign of $C$ indicates the nature of the thermodynamic stability. When $C > 0$, the black hole is locally stable: it cools upon energy loss and can attain thermal equilibrium. When $C < 0$, the configuration is locally unstable, as energy loss leads to an increase in temperature, accelerating the evaporation process.
Using the expressions for $M(r_h)$ and $T(r_h)$ and \eqref{eq:temp}, after some manipulations, we can write

\begin{equation}
    C =
    T\left(\frac{\partial S}{\partial T}\right)_{L,\epsilon}
    =
T\,\left(\frac{\partial S}{\partial r_h}\right)_{L,\epsilon}
\left(\frac{\partial T}{\partial r_h}\right)_{L,\epsilon}^{-1}=
    \frac{
        2\pi r_h^{2}
        \left(2r_h^{2}-3L^{4}\right)
        \left(2r_h^{2}-\epsilon L^{4}\right)
    }{
        -4r_h^{4}
        +(18-2\epsilon)L^{4}r_h^{2}
        -3\epsilon L^{8}
    }.
    \label{eq:heatcap}
\end{equation}

We note that the first two factors appearing after the
second equality sign are positive, namely $T>0$ and
$\left(\frac{\partial S}{\partial r_h}\right)_{L,\epsilon}>0$.
Therefore, the sign of the heat capacity and the locations
of its possible divergences depend solely on the factor
$\left(\frac{\partial T}{\partial r_h}\right)_{L,\epsilon}$. This definition is also consistent with
$C=\frac{du}{dT}
=\mathcal{F}(r_h)
\left(\frac{\partial M}{\partial T}\right)_{L,\epsilon}
=\mathcal{F}(r_h)
\left(\frac{\partial M}{\partial r_h}\right)_{L,\epsilon}
\left(\frac{\partial T}{\partial r_h}\right)_{L,\epsilon}^{-1}$
. Since $\mathcal{F}(r_h)>0$ and
$(\partial M/\partial r_h)_{L,\epsilon}>0$
for $r_h>r_{h,E}$, we confirm that the sign
and possible divergences of the heat capacity
are governed solely by
$(\partial T/\partial r_h)_{L,\epsilon}$.

The heat capacity diverges whenever
\begin{equation}
    -4r_h^{4}+(18-2\epsilon)L^{4}r_h^{2}
    -3\epsilon L^{8}=0,
    \label{eq:davies_condition}
\end{equation}
which identifies a Davies-type turning point
of the temperature.
Solving \eqref{eq:davies_condition} for $r_h^2$ gives the critical
(Davies) radius
\begin{equation}
    r_h^{D}
    =
    \frac{L^{2}}{2}
    \sqrt{(9-\epsilon)+\sqrt{(\epsilon-3)(\epsilon-27)}}.
    \label{eq:davies_radius}
\end{equation}

The branch shown here is selected because
$r_h^{D}>r_{h,E}$. The string contribution modifies the
thermal response by shifting a Davies-type turning point
separating positive- and negative-heat-capacity branches.
The heat capacity is positive for
$r_{h,E}<r_h<r_h^{D}$ and negative for $r_h>r_h^{D}$.
At $r_h=r_h^{D}$, the heat capacity diverges,
indicating a Davies-type critical point separating
the positive- and negative-heat-capacity branches.
However, this divergence alone does not establish
a genuine second-order equilibrium phase transition.

\begin{figure}[h]
\centering
\includegraphics[width=0.6\textwidth]{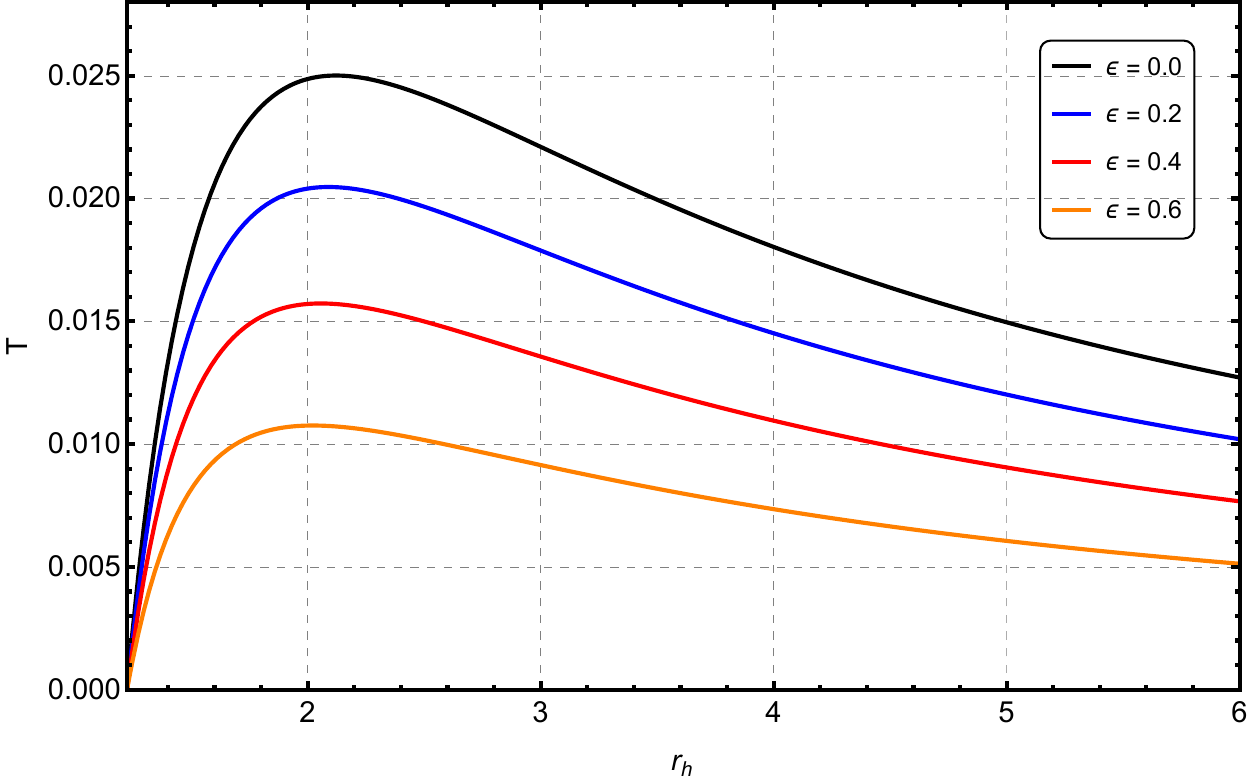}
\caption{Hawking temperature $T$ as a function of $r_h$ 
for $L=1$ and selected values of $\epsilon$.}
\label{fig:temperatura}
\end{figure}

\begin{figure}[h]
\centering
\includegraphics[width=0.6\textwidth]{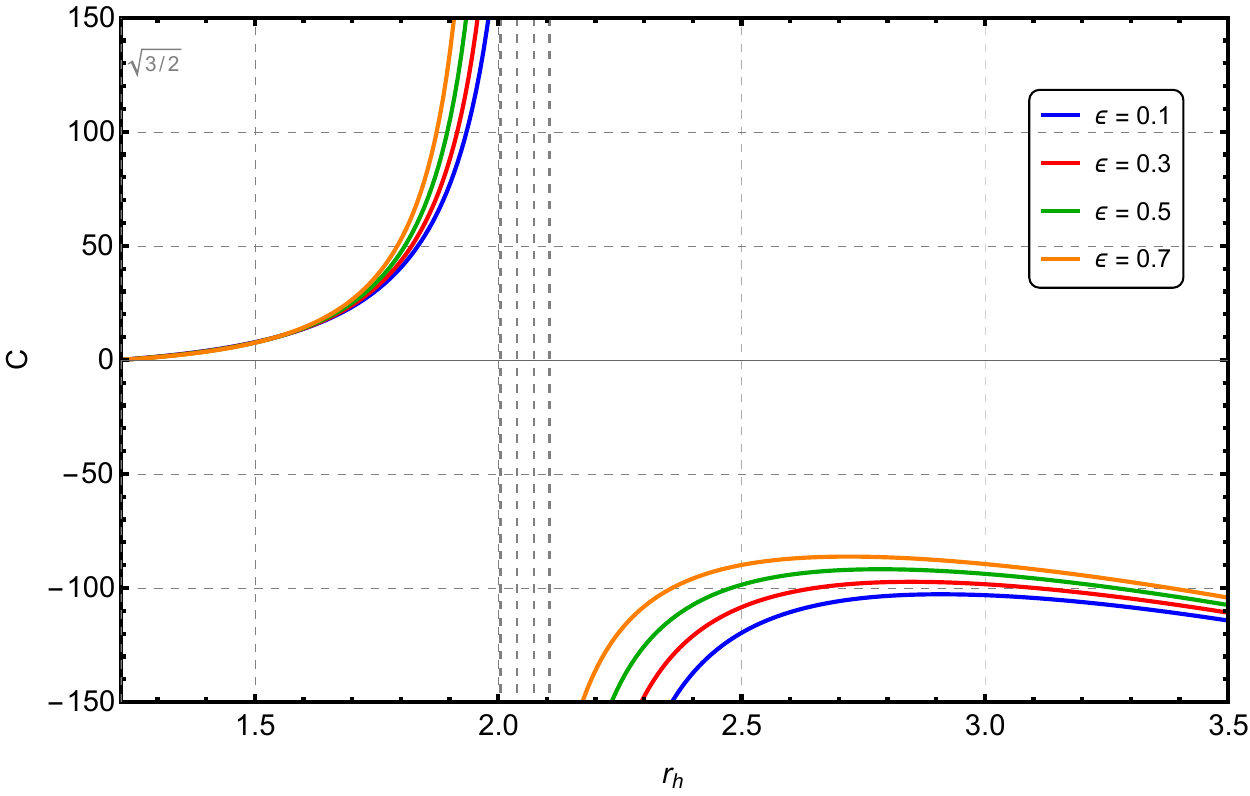}
\caption{Heat capacity $C$ as a function of $r_h$ for 
$L=1$ and selected values of $\epsilon$. }
\label{fig:capacidade}
\end{figure}

The general behavior of the Hawking temperature and
heat capacity is illustrated in
Figs.~\ref{fig:temperatura} and \ref{fig:capacidade}.
As discussed above, the sign of the heat capacity
is governed by the derivative
$(\partial T/\partial r_h)_{L,\epsilon}$.
Following the interpretation of the radial evolution
discussed in Ref. \cite{Estrada:2023cyx}, we may consider a hypothetical process in which the horizon radius
progressively decreases, corresponding to moving
from right to left in Fig.~\ref{fig:temperatura}.
For $r_h>r_h^D$, the heat capacity is negative,
$C<0$, indicating that the black hole becomes hotter
as it loses energy and its horizon contracts.
At $r_h=r_h^D$, the heat capacity diverges,
marking a Davies-type turning point that separates
two branches with different local thermodynamic
stability properties.
For $r_{h,E}<r_h<r_h^D$, the heat capacity becomes
positive, $C>0$, so that a further decrease in the
horizon radius is accompanied by a decrease in
temperature, while the black hole continues to lose energy.
As the extremal radius $r_{h,E}$ is approached,
both the Hawking temperature and the heat capacity
tend to zero.
This behavior is consistent with interpreting
the extremal configuration as a possible
black-hole remnant, understood as a limiting state
that could remain after radial contraction ceases
\cite{Adler:2001vs}.
Nevertheless, the vanishing of the temperature and
heat capacity does not guarantee that evaporation
necessarily terminates, since establishing this
would require a detailed analysis of the evaporation
process beyond the scope of the present work.

Two limiting cases provide useful consistency checks.
In the limit $\epsilon\to0$, Eq.~\eqref{eq:heatcap}
reduces to the Hayward expression. In the further
limit $L\to0$, the regularisation is removed and
the heat capacity reduces to $C=-2\pi r_h^2<0$,
coinciding with the Schwarzschild result, although
the string-cloud contribution remains present
for $\epsilon\neq0$.

In summary, the thermodynamic analysis identifies the Davies radius, ($r_h^{D}$), which signals a potential change in thermodynamic stability. The dependence of $T$ and $C$ on $\epsilon$ is illustrated in Figs.~\ref{fig:temperatura} and \ref{fig:capacidade}. In this way, the screening factor introduced through the MGD algorithm not only generates a regular string-fluid geometry with nontrivial behavior at both short and long distances, but also significantly modifies its thermodynamic evolution. In particular, whereas the usual cloud-of-strings geometry is thermodynamically unstable, with a specific heat $C=-8\pi M^2(1-\epsilon)^{-2}$, the regularization proposed here introduces a Davies-type transition between thermodynamic branches with opposite signs of the heat capacity. Indeed, as can be seen in Fig.\ref{fig:temperatura}, from right to left, at a critical radius $r_h^{D}$ the black hole changes from an unstable branch to a stable one. From this point onward, the temperature decreases until it vanishes at the extremal radius, leading to the formation of a black hole remnant. It is worth emphasizing that, for fixed $M$, the extremal radius depends explicitly on the string-cloud parameter $\epsilon$. Consequently, depending on its value, the remnant may or may not be of Planckian size, exhibiting a behavior compatible with that expected in LQG-inspired scenarios.

\section{Quasinormal modes}\label{sec:Quasinormal_modes}
Quasinormal modes (QNMs) provide a key description of the dynamical response of black holes under perturbations. These modes are characterized by damped oscillations with complex-valued frequencies: the real part determines the oscillation rate, whereas the imaginary part controls the damping due to energy loss. In this subsection, we investigate the scalar QNM spectrum of the regular fluid of string spacetime by employing the WKB approach through the implementation \cite{Konoplya:2019hlu}.
Consider a scalar field $\Psi$ of mass $\mu$%
\footnote{{The mass term is retained for completeness in the derivation, while all numerical results presented below correspond to a massless scalar field, $\mu=0$.}}
evolving in a background geometry described by the metric tensor $g_{\mu\nu}$. Its dynamics are governed by the Klein--Gordon equation
\begin{equation}
\frac{1}{\sqrt{-g}} \, \partial_\nu \left( g^{\mu\nu} \sqrt{-g} \, \partial_\mu \Psi \right) - \mu^2 \Psi = 0.
\label{42}
\end{equation}

We analyze a {scalar perturbation} propagating in the static, spherically symmetric geometry defined by the line element (\ref{MGD1}). For this metric, the determinant satisfies $\sqrt{-g} = r^2 \sin\theta$. After substituting this expression into the Klein--Gordon equation and expanding the derivatives explicitly, one finds
\begin{widetext}
    \begin{equation}
 -\frac{1}{f(r)} \frac{\partial^2 \Psi}{\partial t^2}
+ \frac{1}{r^2} \frac{\partial}{\partial r}\!\left(r^2 f(r) \frac{\partial \Psi}{\partial r}\right) 
+ \frac{1}{r^2 \sin\theta}
\frac{\partial}{\partial \theta}\!\left(\sin\theta \frac{\partial \Psi}{\partial \theta}\right)
+ \frac{1}{r^2 \sin^2\theta}\frac{\partial^2 \Psi}{\partial \phi^2}
- \mu^2 \Psi = 0 .
\end{equation}
\end{widetext}

Taking advantage of the spherical symmetry, we decompose the scalar field using the ansatz
\begin{equation}
\Psi(t,r,\theta,\phi) =
e^{-i\omega t}
R(r)
Y_{\nu m}(\theta,\phi),
\end{equation}
where $Y_{\nu m}$ denote the spherical harmonics. As discussed in the previous section, the normalization
of the Killing vector also affects the quasinormal
frequencies. Under the temporal coordinate redefinition
$\widehat{t}=\sqrt{1-\epsilon}\,t$, the scalar field
perturbation can be expressed as
\begin{equation}
\Psi(\widehat{t},r,\theta,\phi)
\sim e^{-i\widehat{\omega}\widehat{t}}
\,\,,\,\,
\widehat{\omega}
=\frac{\omega}{\sqrt{1-\epsilon}}.
\end{equation}
This relation applies to both the real and imaginary
parts of the quasinormal frequencies.
For consistency, throughout this section we retain
the original coordinate time $t$ and report the
frequencies $\omega$, which differ from
$\widehat{\omega}$ by the constant factor
$\sqrt{1-\epsilon}$.

Introducing this decomposition into the Klein--Gordon equation and defining the tortoise coordinate $r_*$ by
\begin{equation}
\frac{dr_*}{dr} = \frac{1}{f(r)},
\end{equation}
together with the redefinition
\begin{equation}
R(r) = \frac{u(r)}{r},
\end{equation}
the radial sector assumes a Schrödinger-type form,
\begin{equation}
\frac{d^2 u}{dr_*^2}
+
\left[
\omega^2 - V(r)
\right] u = 0.
\end{equation}

The corresponding effective potential reads
\begin{equation}
V(r)
=
f(r)
\left[
\frac{\nu(\nu+1)}{r^2}
+
\frac{f'(r)}{r}
+
\mu^2
\right]
\label{potential}.
\end{equation}

The effective potential incorporates the contributions from the spacetime geometry, the angular momentum barrier, and the scalar field mass. Consequently, it governs the propagation of scalar waves and determines the quasinormal spectrum of the system. To compute the QNMs, we employ the WKB approximation, a widely used semi-analytical method in black hole perturbation theory. This technique treats the radial equation as an effective quantum-mechanical scattering problem and approximates the solution around the maximum of the potential barrier. From this procedure, one obtains the complex frequencies associated with damped oscillatory modes, with particularly good accuracy for low overtone numbers.

In the present analysis, we use the sixth-order WKB formalism \cite{Konoplya:2003ii}. Within this framework, the quasinormal frequencies are determined from the effective potential (see Eq.~\ref{potential}) via
\begin{equation} \label{FormulaQNM}
    {i}
    \frac{\omega^2 - V_0}{\sqrt{-2V_0^{''}}}
    - \Lambda_2 - \Lambda_3 - \Lambda_4
    - \Lambda_5 - \Lambda_6
    = n + \frac{1}{2},
\end{equation}
where $V_0$ is the peak value of the effective potential and $V_0^{''}$ is its second derivative evaluated at the peak and where $n=0,1,2,\ldots$ is the overtone number, with $n=0$ corresponding to the fundamental quasinormal mode. The correction terms $\Lambda_i$, for $2 \leq i \leq 6$, are explicitly given in Refs.~\cite{Iyer:1986np,Konoplya:2003ii}.

The computational procedure can be summarized as follows:
\begin{enumerate}

    \item Determine the location of the maximum of the effective potential $V(r)$ and evaluate the required derivatives at that point, since the WKB expansion is centered around this maximum.

    \item Insert these quantities into the sixth-order WKB formula, which connects the complex frequency $\omega$ with the value and derivatives of the potential at its maximum.

    \item Solve the resulting algebraic equation for $\omega$ to obtain the quasinormal frequencies corresponding to a chosen overtone number $n$.

    \item Repeat the analysis for different model parameters to investigate how the quasinormal spectrum changes.
\end{enumerate}

\begin{figure}[h]
\centering
\includegraphics[scale=1]{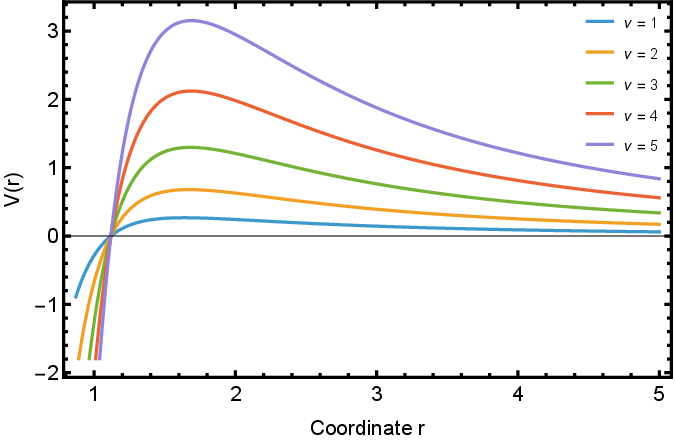} 
\caption{Effective potential $V(r)$ for a massless scalar field as a function of the radial coordinate $r$, considering the fundamental mode and different values of the parameter $\nu$ with $L=1/2$, $\epsilon=0.1$ and $M=0.515625$. The height of the potential barrier increases monotonically with $\nu$,  while its peak is shifted slightly toward larger values of $r$.}
\label{efetivo}
\end{figure}

\begin{figure}[ht]
\centering
\includegraphics[scale=0.5]{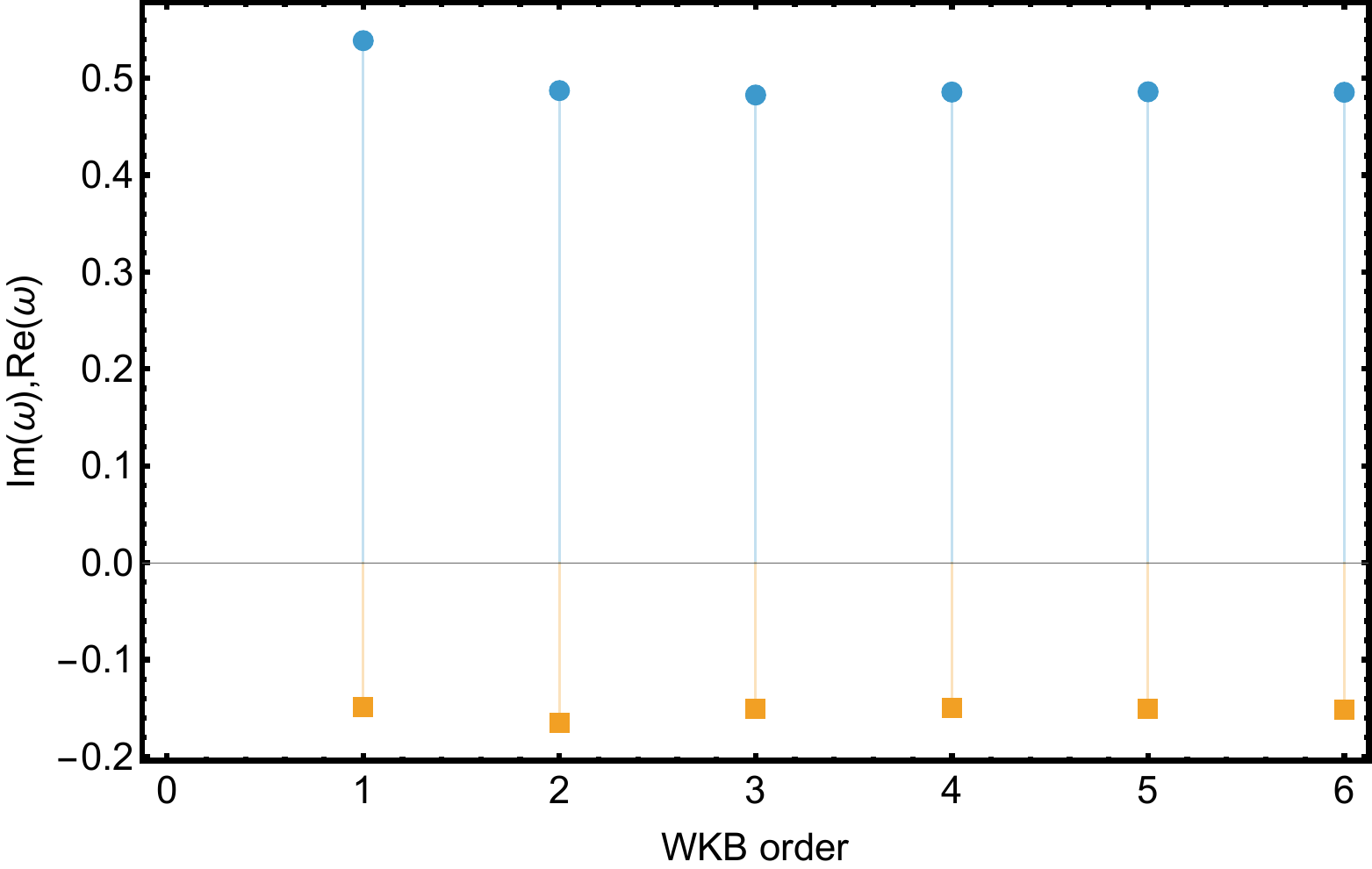} 
\caption{Dependence of the quasinormal frequencies on the WKB approximation order for the fundamental mode with $\nu=1$. The real and imaginary parts rapidly converge as the WKB order increases, with the frequencies becoming nearly unchanged from the third order onward.}
\label{convergence}
\end{figure}

\begin{figure}[ht]
\centering
\includegraphics[scale=1]{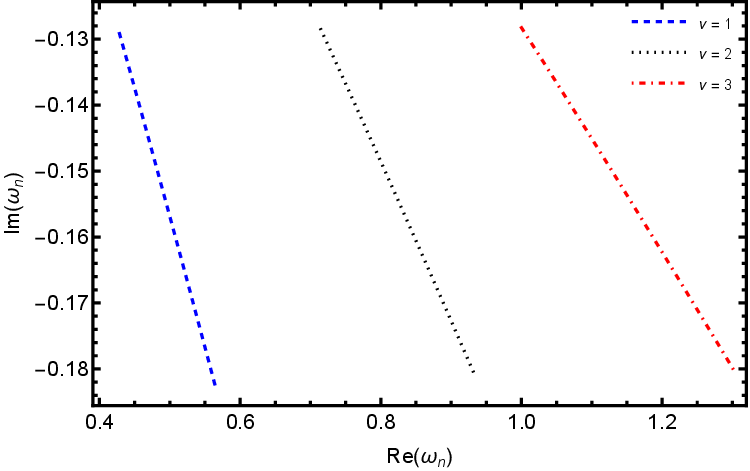} 
\caption{Quasinormal frequencies in the complex plane for $n=0$, obtained by varying $\epsilon$ from $0.01$ to $0.17$ with a step of $0.01$. The calculations were performed with $L=1/2$ and $M=0.515625$ for $\nu=1$, $2$, and $3$.   
}
\label{n0}
\end{figure}

\begin{figure}[ht]
\centering
\includegraphics[scale=1]{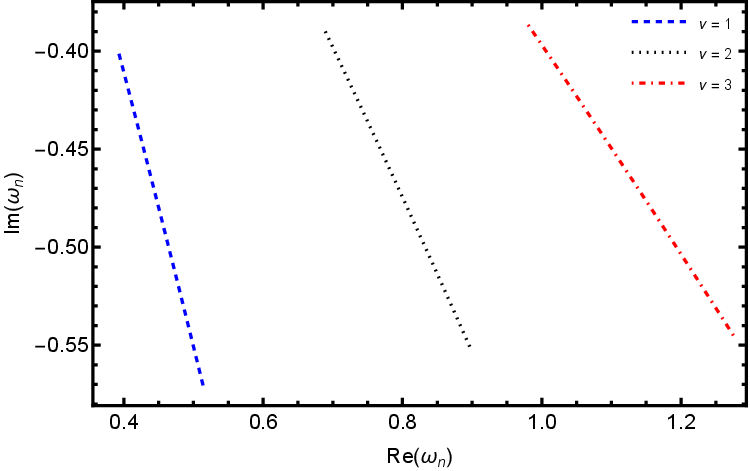} 
\caption{Quasinormal frequencies in the complex plane for the first overtone ($n=1$), computed by varying $\epsilon$ from $0.01$ to $0.17$. The parameters were fixed at $L=1/2$ and $M=0.515625$, while $\nu=1$, $2$, and $3$.}
\label{n1}
\end{figure}

\begin{figure}[ht]
\centering
\includegraphics[scale=1]{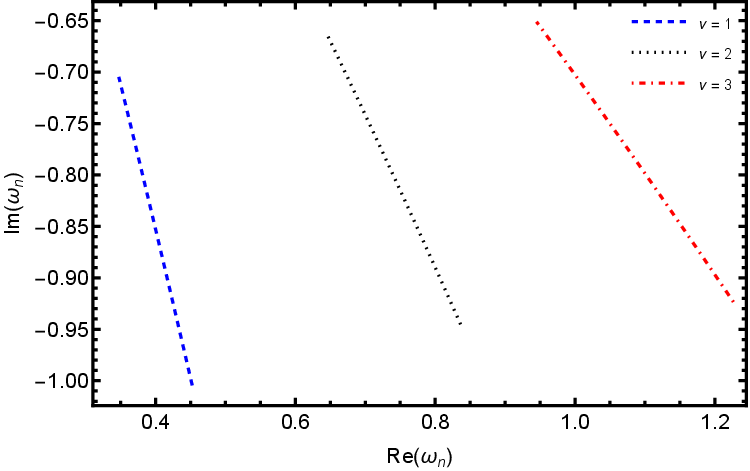} 
\caption{Quasinormal frequencies in the complex plane for the second overtone ($n=2$), with $\epsilon$ ranging from $0.01$ to $0.17$. The results correspond to $L=1/2$, $M=0.515625$, and $\nu=1$, $2$, and $3$.   
}
\label{n2}
\end{figure}

Having established the scalar perturbation equation and the sixth-order WKB formalism, we now present the QNM results. It is instructive to examine the behavior of the effective potential, since its shape largely determines the oscillation frequencies and damping rates. Figure~\ref{efetivo} shows the effective potential $V(r)$ for the massless scalar field $\mu=0$ with overtone number $n=0$ and several values of the parameter $\nu$. In all cases, the potential exhibits the characteristic barrier-like profile required for the application of the WKB approximation. As $\nu$ increases, the height of the potential barrier grows significantly, while the position of its maximum is shifted slightly toward larger radial coordinates. The increase in the peak is monotonic, indicating that larger values of $\nu$ produce a stronger effective confinement of the scalar perturbations around the black hole. In contrast, the overall width and shape of the barrier undergo only moderate changes. Using the same notation as in \cite{Lin:2013ofa}, let us consider the mass terms of $L$ in the form $M=(1+L^4/2)/2$ . 
Figure~\ref{convergence} illustrates the dependence of the quasinormal frequencies on the WKB order for the fundamental mode with $\nu=1$. Both the real and imaginary parts exhibit rapid convergence as the WKB order increases. While the first-order approximation shows the largest deviation, the frequencies stabilize from the third order onward, with only negligible variations between successive orders. 
Figures~\ref{n0}--\ref{n2} present the quasinormal frequencies in the complex plane for the first three overtones, corresponding to $n=0$, $n=1$, and $n=2$, respectively. The frequencies were computed by varying the parameter $\epsilon$ from $0.01$ to $0.17$ in steps of $0.01$, while keeping $L=1/2$ and $M=0.515625$ fixed. For all values of $\nu$, the modes evolve along nearly straight lines in the complex-frequency plane as $\epsilon$ increases. Moreover, at fixed $\epsilon$, increasing $\nu$ shifts the frequencies
toward larger values of $\mathrm{Re}(\omega_n)$ and smaller values of
$|\mathrm{Im}(\omega_n)|$, corresponding to higher oscillation
frequencies and slower damping. As expected, higher overtones exhibit substantially larger damping rates, whereas the real part of the frequency changes comparatively less.


\subsection{Estimation of uncertainties and reliability of the WKB approximation}
\label{subsec:WKB_uncertainty}
To evaluate the uncertainties associated with the quasinormal frequencies shown in Figs.~\ref{n0}--\ref{n2}, we consider the truncation of the sixth-order WKB approximation, whose accuracy may decrease for higher overtone numbers relative to the multipolar number $\nu$. For this purpose, we employ the uncertainty estimator proposed by Konoplya et al.~\cite{Konoplya:2019hlu}, defined as
\begin{equation}
\label{WKB_error}
\Delta_k=
\frac{|\omega_{k+1}-\omega_{k-1}|}{2},
\end{equation}
where $\omega_k$ represents the quasinormal frequency calculated using the WKB approximation of order $k$. This estimator compares approximations separated by two orders, taking into account that successive WKB corrections contribute alternately to the real and imaginary parts of $\omega^2$.In our case, the frequencies presented above correspond to the sixth-order approximation and were obtained using Eq.~\eqref{FormulaQNM}. To determine the fifth- and seventh-order frequencies, we modify the truncation of that expression, obtaining, respectively,
\begin{equation}
\label{WKB_order5}
i\frac{\omega_5^2-V_0}{\sqrt{-2V_0^{''}}}
-\sum_{j=2}^{5}\Lambda_j
=n+\frac{1}{2},
\end{equation}
and
\begin{equation}
\label{WKB_order7}
i\frac{\omega_7^2-V_0}{\sqrt{-2V_0^{''}}}
-\sum_{j=2}^{7}\Lambda_j
=n+\frac{1}{2}.
\end{equation}

Here, the corrections $\Lambda_j$ are understood to be evaluated at the corresponding order, following the higher-order extension of the WKB formalism developed by Konoplya and collaborators~\cite{Konoplya:2019hlu}. In particular, the seventh-order approximation requires the inclusion of the additional correction $\Lambda_7$.

To calculate these frequencies, we reproduced Konoplya's procedure through a mathematically equivalent implementation based on the WKB expansion around the maximum of the effective potential. In particular, we used the effective potential defined in Eq.~\eqref{potential}, considering a massless scalar field, $\mu=0$, and evaluating its derivatives with respect to the tortoise coordinate $r_*$, according to $\frac{d}{dr_*}=f(r)\frac{d}{dr}$. For each configuration $(\nu,n)$, the physical parameters were kept fixed, so that the differences between $\omega_5$, $\omega_6$, and $\omega_7$ arise exclusively from the truncation order of the WKB expansion. Applying Eq.~\eqref{WKB_error} to the sixth-order approximation, we obtain
\begin{equation}
\label{WKB_error6}
\Delta_6=
\frac{|\omega_7-\omega_5|}{2}.
\end{equation}
We also introduce the percentage relative uncertainty
\begin{equation}
\label{WKB_relative_error}
\delta_6=
\frac{\Delta_6}{|\omega_6|}\times100\%,
\end{equation}
which allows us to compare the estimated uncertainties for quasinormal modes of different magnitudes. It is important to note that $\Delta_6$ represents an uncertainty estimator associated with the truncation of the WKB expansion, rather than a rigorous bound on the error relative to the exact quasinormal frequency.
First, we consider $\epsilon=0.10$, keeping the same parameters used in the previous analysis, $L=1/2$, $M=0.515625$, and $\mu=0$. For each combination $(\nu,n)$, we calculate the fifth-, sixth-, and seventh-order frequencies and evaluate the uncertainties using Eqs.~\eqref{WKB_error6} and \eqref{WKB_relative_error}. The results are presented in Table~\ref{tab:WKB_errors}.
\begin{table*}[htbp]
\centering
\caption{Quasinormal frequencies obtained using the fifth-, sixth-, and seventh-order WKB approximations, together with their uncertainty estimators for $\epsilon=0.10$, considering $L=1/2$, $M=0.515625$, and $\mu=0$.}
\label{tab:WKB_errors}
\small
\setlength{\tabcolsep}{4pt}
\begin{tabular}{ccccccc}
\hline\hline
$\nu$ & $n$ & $\omega_5$ & $\omega_6$ & $\omega_7$ & $\Delta_6$ & $\delta_6$ (\%) \\
\hline
1 & 0 & $0.486162-0.150959i$ & $0.485635-0.151123i$ & $0.485524-0.150766i$ & $0.000334$ & $0.0656$ \\
1 & 1 & $0.445431-0.470242i$ & $0.444339-0.471398i$ & $0.443293-0.470412i$ & $0.001073$ & $0.1656$ \\
1 & 2 & $0.391030-0.831776i$ & $0.392221-0.829248i$ & $0.390210-0.828299i$ & $0.001786$ & $0.1947$ \\
2 & 0 & $0.805159-0.149651i$ & $0.805113-0.149660i$ & $0.805109-0.149641i$ & $0.000025$ & $0.0031$ \\
2 & 1 & $0.776932-0.455924i$ & $0.776773-0.456017i$ & $0.776724-0.455934i$ & $0.000104$ & $0.0116$ \\
2 & 2 & $0.727774-0.781358i$ & $0.727798-0.781333i$ & $0.727721-0.781261i$ & $0.000055$ & $0.0052$ \\
3 & 0 & $1.125549-0.149320i$ & $1.125540-0.149321i$ & $1.125540-0.149319i$ & $4.373006\times10^{-6}$ & $0.0004$ \\
3 & 1 & $1.104625-0.451603i$ & $1.104592-0.451616i$ & $1.104587-0.451605i$ & $0.000019$ & $0.0016$ \\
3 & 2 & $1.065571-0.764358i$ & $1.065569-0.764359i$ & $1.065564-0.764352i$ & $4.474998\times10^{-6}$ & $0.0003$ \\
\hline\hline
\end{tabular}
\end{table*}

The results in Table~\ref{tab:WKB_errors} show relatively small differences among the frequencies obtained at fifth, sixth, and seventh order. However, these differences depend on the multipolar and overtone numbers. For $\nu=1$, the estimated relative uncertainty increases progressively from the fundamental mode ($n=0$) to the second overtone ($n=2$). In contrast, for $\nu=3$, the uncertainties are considerably smaller for all three modes analyzed, indicating greater stability of the WKB approximation as the multipolar number increases. This behavior is consistent with the greater reliability of the WKB method when the multipolar number is sufficiently large compared with the overtone number ($\nu\gg n$). Moreover, higher overtones exhibit faster damping, associated with the increase in $|\operatorname{Im}(\omega_6)|$, although their calculation using the WKB method requires greater caution due to the possible loss of accuracy.

To extend this evaluation to all configurations considered in Figs.~\ref{n0}--\ref{n2}, we repeat the procedure for values of the string parameter ranging from $\epsilon=0.01$ to $\epsilon=0.17$, in steps of $0.01$, while keeping $L=1/2$ and $M=0.515625$ fixed. For each combination $(\nu,n)$, we determine the minimum and maximum values of the relative uncertainty $\delta_6$ over 17 discrete values of $\epsilon$. The results are presented in Table~\ref{tab:WKB_error_ranges}.
\begin{table}[htbp]
\centering
\caption{Ranges of the sixth-order WKB relative uncertainty for $0.01\leq\epsilon\leq0.17$, considering $L=1/2$, $M=0.515625$, and $\mu=0$.}
\label{tab:WKB_error_ranges}
\begin{tabular}{cccc}
\hline\hline
$\nu$ & $n$ & $\delta_{6,\min}$ (\%) & $\delta_{6,\max}$ (\%) \\
\hline
1 & 0 & 0.0479 & 0.0955 \\
1 & 1 & 0.1194 & 0.2415 \\
1 & 2 & 0.1626 & 0.2512 \\
2 & 0 & 0.0021 & 0.0049 \\
2 & 1 & 0.0076 & 0.0191 \\
2 & 2 & 0.0043 & 0.0133 \\
3 & 0 & 0.0003 & 0.0006 \\
3 & 1 & 0.0010 & 0.0027 \\
3 & 2 & 0.0002 & 0.0018 \\
\hline\hline
\end{tabular}
\end{table}

Table~\ref{tab:WKB_error_ranges} shows that the estimated uncertainties vary with $\epsilon$ due to its influence on the effective potential. The largest values correspond to $(\nu,n)=(1,2)$, whereas for $\nu=3$ they are considerably smaller. However, these small estimators do not guarantee the accuracy of the frequencies, since differences between different WKB orders reflect stability with respect to truncation, but not necessarily proximity to the exact solution. A more rigorous validation using independent numerical methods or Padé approximants would require an additional analysis beyond the scope of the present work.

It is worth noting that, as pointed out in Ref. \cite{Konoplya:2019hlu}, adding higher-order terms to the WKB expansion may initially improve the approximation, but not necessarily indefinitely. Therefore, increasing the approximation order does not guarantee greater accuracy in the resulting frequencies. Furthermore, the standard WKB approximation is generally more reliable for low overtones, particularly when $\nu \gg n$, whereas modes with $n$ comparable to or greater than $\nu$ require greater caution. Consequently, our results for higher overtones, especially those corresponding to $(\nu,n)=(1,2)$, should be interpreted with caution, since the uncertainty estimates reflect the sensitivity to the truncation order of the WKB expansion and do not constitute an exact error bound for the calculated frequencies


\section{Discussion and conclusions}\label{sec:Discussion_and_conclusions}

In this work, we have investigated whether black holes supported by a fluid of strings can be consistently regularized without destroying the characteristic large-distance behavior of the string sector. In the present work, the string fluid is characterized by the equation of state $(p_t)_T=W(r)\rho_T$, where the function $W(r)$ is obtained from the gravitational decoupling procedure discussed in the main text. By construction, $\displaystyle \lim_{r\rightarrow\infty}W(r)=0$, so that 
$\displaystyle \lim_{r\rightarrow\infty}(p_t)_T=0$, and the matter source asymptotically reduces to the standard pressureless cloud-of-strings solution of Ref.~\cite{Letelier:1979ej}. Unlike conventional regular black holes, where replacing the point mass by an effective mass function is sufficient to remove the central singularity, we have shown that this standard mechanism fails in the presence of the cloud-of-strings contribution. This result highlights that the string sector plays an independent role in the ultraviolet structure of the spacetime and therefore requires a different regularization mechanism. Motivated by this observation, we have constructed a regular black hole supported by an effective anisotropic string fluid, providing a spacetime that remains regular at short distances while recovering the characteristic cloud-of-strings behavior in the infrared regime.  

A key outcome of this construction is that the string sector modifies the spacetime at both short and large distances in a physically consistent manner. Near the origin, the additional string contribution introduces subleading corrections to the Hayward core while preserving its leading de Sitter behavior and the regularity of the curvature invariants. At large distances, the solution asymptotically recovers the cloud-of-strings geometry, but also develops an additional correction proportional to $\epsilon r^{-3}$, whereas the usual LQG Hayward correction behaves as $r^{-4}$. Consequently, the string sector introduces a longer-range modification of the geometry that could, in principle, become relevant before the standard Hayward correction becomes appreciable. These results indicate that the proposed regularization does not merely remove the central singularity, but also modifies the spacetime across different physical scales while maintaining a consistent effective string-fluid interpretation.

The geometric analysis further shows that the regularization mechanism preserves the existence of the characteristic black-hole configurations, including non-extremal and extremal solutions, while admitting a regular horizonless compact object beyond the critical configuration. At the same time, the string sector modifies the internal
geometry through subleading deformations
of the near-core spatial metric. Although these corrections
do not modify the leading local $S^3$ spatial geometry,
they introduce a nontrivial subleading contribution to
the three-dimensional Ricci scalar, while preserving
the leading de Sitter behavior near the origin.
However, these local geometric modifications do not,
by themselves, establish a change in global topology.

The physical consequences of the proposed regularization
extend beyond the geometric sector. From the thermodynamic
perspective, the string contribution
modifies the thermal response by shifting
a Davies-type turning point separating positive- and
negative-heat-capacity branches, in contrast with the
standard cloud-of-strings geometry, whose heat capacity
remains always negative.
Due to the presence of matter sources
in the energy-momentum tensor, the thermodynamic relation
in the present model takes a modified form. Accordingly,
we introduce a local energy variation satisfying
$du=T\,dS$. The heat capacity is then consistently
evaluated from this relation at fixed $L$ and $\epsilon$.
Moreover, the Hawking temperature vanishes at a finite
extremal radius, identifying a possible
remnant configuration. At fixed $L$, the extremal radius
is independent of $\epsilon$, whereas the corresponding
extremal mass depends on the string-cloud parameter.

Beyond thermodynamics, the quasinormal-mode analysis shows that the modifications introduced by the string sector are also encoded in the dynamical response of the spacetime. The quasinormal frequencies vary systematically with $\epsilon$, affecting both the oscillation frequencies and the damping rates of scalar perturbations. Together, these results indicate that the string sector leaves observable signatures not only on the equilibrium properties of the black hole but also on its dynamical evolution under external perturbations.

Overall, our results show that regularizing string-supported black holes requires going beyond the standard regular-black-hole paradigm based solely on an effective mass function. Instead, the present construction demonstrates that the string sector itself plays an active role in determining the ultraviolet structure of the spacetime, while preserving the characteristic infrared behavior of cloud-of-strings geometries. This framework provides a physically consistent realization of a regular string-fluid black hole in which the matter source, geometry, thermodynamic evolution, and dynamical response are intrinsically interconnected. We hope that these results will motivate further investigations of regular black holes supported by physically motivated matter sources, particularly in scenarios where quantum-gravity-inspired effects coexist with nontrivial matter sectors.

\section*{Acknowledgements}\label{Acknowledgements}
LCNS would like to thank Conselho Nacional de Desenvolvimento Científico e Tecnológico~-~Brazil (CNPq) for financial support under Research Project No. 443769/2024-9 and Research Fellowship No. 314815/2025-2. L.G.B. acknowledges the financial support of the Coordenação de Aperfeiçoamento de Pessoal de Nível Superior (CAPES), Brazil (Finance Code 001).

\section*{Data Availability Statement}

Data sharing is not applicable to this article as no new data were created or analyzed in this study.

\appendix

\section{Brief Review of the Hayward–Planck Star Model}
\label{ApendicePlanckStar}

We employ the following generalization of the energy density of the Planck-star model or Hayward \cite{DeLorenzo:2014pta,Hayward:2005gi},
\begin{equation}
 \rho_{H}
 =
 \frac{1}{4\pi/3}
 \frac{M^2}
 {\left(L^2M+\dfrac{r^3}{L^2}\right)^2},
\end{equation}
\begin{equation}
    (p_t)_H =-\frac{3L^4M^2\left(L^4M-2r^3\right)}
{4 \pi\left(L^4M+r^3\right)^3}
\end{equation}
where the units are chosen such that $[M]^{-1}=[L]=\ell_p$. It is straightforward to note that one can define an equation of state of the form $(p_t)_H=\omega_0(r)\,\rho_H$, where

\begin{equation}\label{omega0}
\omega_0(r)
=
\frac{
-\,LM
+
2\left(\dfrac{r}{L}\right)^3
}{
LM
+
\left(\dfrac{r}{L}\right)^3
}.
\end{equation}

The corresponding mass function is
\begin{equation}
m(r)
=
4\pi\int \rho(r)\,r^2\,dr
=
-\frac{M^2L^4}{L^4M+r^3}+M
=
\frac{Mr^3}
{L^3\left(LM+\dfrac{r^3}{L^3}\right)}.
\end{equation}

The Hayward-type solution \cite{Hayward:2005gi} is given by
\begin{equation}
f(r)
=
1-\frac{2G\,m(r)}{r}.
\end{equation}

Setting $G=\ell_p^2$, we obtain
\begin{equation}
f(r)
=
1-\frac{2M}{r}\cdot
\frac{r^3}
{L^3\left(LM+\dfrac{r^3}{L^3}\right)}=1-\frac{2M}{r}\cdot s(r) \label{EqSF}
\end{equation}
where $s(r)$ denotes the screening factor.

At large scale, for $x=M/r\ll 1$ (equivalently, $r\gg M$), one obtains
\begin{equation}
\label{OmegaCeroLargaEscala}
\omega_0(x)
\approx
2+ \mathcal{O}(x^3)
=
2
+ \mathcal{O}\!\left(\frac{M^3}{r^3}\right).
\end{equation}

It is straightforward to verify that, in this regime, the Hayward energy density behaves as
\begin{align}
\label{DensidadCeroLargaEscala}
\rho_H(x)
&\approx
\frac{3L^4}{4\pi M^4}\,x^6
+\mathcal{O}(x^9)
=
\frac{3L^4M^2}{4\pi r^6}
+\mathcal{O}(r^{-9})
\approx \mathcal{O}(x^6).
\end{align}

Thus, substituting Eqs.~\eqref{DensidadCeroLargaEscala} and \eqref{OmegaCeroLargaEscala} into Eq.~\eqref{TangencialCero}, we obtain
\begin{align}
\label{TangencialCeroLargaEscala}
(p_t)_0
&\approx
\left(
2-3\frac{L^4}{M^2}x^3
+\mathcal{O}(x^6)
\right)
\left(
\frac{3L^4}{4\pi M^4}x^6
+\mathcal{O}(x^9)
\right)\approx
\frac{3L^4}{2\pi M^4}x^6
+\mathcal{O}(x^9)
\approx \mathcal{O}(x^6).
\end{align}

At this regime we obtain
\begin{align}
s(r)
&=\frac{1}{1+\dfrac{L^4}{M^2}x^3} \approx 1-\frac{L^4}{M^2}x^3+\mathcal{O}(x^6)
=
1-\frac{ML^4}{r^3}
+\mathcal{O}(r^{-6})
\label{SFLargaEscala}
\\
\Rightarrow \quad
f(r)
&\approx
1-\frac{2M}{r}
+\frac{2L^4M^2}{r^4}
+\mathcal{O}(r^{-7})
\label{LQG}
\end{align}

The last term may be interpreted as an effective quantum-gravitational repulsive correction at large scales. The terms of order $x^6$ indicate that the quantum corrections to the Hayward seed density become negligible at large scales. Nevertheless, they may still remain marginally testable up to a few characteristic length scales before the $r^{-1}$ contribution to the metric tensor \eqref{LQG} becomes dominant. This is consistent with the expectation that effects arising from Loop Quantum Gravity are generally more pronounced at short distances, as discussed in the Introduction.

At short scales $x\ll 1$ (equivalently, $r\ll L$), one obtains
\begin{align} \label{OmegaCeroCortaH}
\omega_0(x)
&=
\frac{
-1+2x^3/(LM)
}{
1+x^3/(LM)
}
\approx -1 +\mathcal{O}(x^3)
 \approx -1 +
\mathcal{O}\!\left(\frac{r^3}{L^3}\right)
\end{align}
which resembles a de Sitter equation of state. It is straightforward to verify that, in this regime, the Hayward energy density behaves as
\begin{align} \label{DensidadCeroCorta}
\rho_H(x)
&=
\frac{3}{4\pi L^4}
\frac{1}{\left(1+\dfrac{x^3}{LM}\right)^2}
 \approx \frac{3}{4\pi L^4} +
\mathcal{O}(x^3) = \frac{3}{4\pi L^4}+\mathcal{O}\!\left(\frac{r^3}{L^{3}}\right).
\end{align}

Thus, substituting Eqs.~\eqref{DensidadCeroCorta} and
\eqref{OmegaCeroCortaH} into the equation of state Eq.~\eqref{TangencialCero}, we obtain
\begin{align} \label{TangencialCeroCorta}
(p_t)_0
& \approx
-\frac{3}{4\pi L^4} +
\mathcal{O}\!\left(\frac{r^3}{L^{3}}\right).
\end{align}

Therefore, at short distances, the seed sector approaches a de Sitter vacuum,
characterized by the equation of state
\begin{equation} \label{DeSitterCorta}
(p_t)_0\simeq -\rho_H,
\end{equation}
which is the expected behavior of the Hayward core. At this regime one finds
\begin{align}
    &s(r)=\frac{x^3}{x^3+L\,M} \approx \frac{x^3}{L\,M}+\mathcal{O}(x^6) \approx \frac{r^3}{L^4 \, M} \nonumber \\
    & \Rightarrow f(r) \approx 1- \frac{2}{L^4} r^2 \label{dSCore}
\end{align}

Thus, the geometry approaches a de Sitter core at short distances.

\section{A Brief Overview of the Cloud-of-Strings Model}\label{sec:A_Brief_Overview_of_the_Cloud-of-Strings_Model}

In this appendix, we summarize the main ingredients of the cloud-of-strings model introduced in Ref.~\cite{Letelier:1979ej}, which are required for the analysis developed in this work. A relativistic string (1-brane) is described by two internal coordinates $\lambda^a=(\lambda^0,\lambda^1)$ and is embedded in spacetime through the mapping $x^\mu=x^\mu(\lambda)$. As the string evolves, it sweeps out a two-dimensional timelike worldsheet. The dynamics of such an object are governed by the Nambu--Goto action, which is proportional to the area of the worldsheet,

\begin{equation} \label{AccionNambuGoto}
    S_{NG}= \int \mathcal{M} \, dA = \int \mathcal{M} \sqrt{-h} \, d\lambda^0 d\lambda^1 \, , \qquad
    h_{ab}=g_{\mu\nu}
    \frac{\partial x^\mu}{\partial \lambda^a}
    \frac{\partial x^\nu}{\partial \lambda^b}.
\end{equation}

The constant $\mathcal{M}$ is chosen so that the action remains dimensionless in natural units. Typically,
$[\mathcal{M}]=[T_0/c]=[\mathrm{Force}]=[\mathrm{Energy}/L]=[L^{-2}]$,
where $T_0$ denotes the string tension and $c$ is the speed of light. The tensor $h_{ab}$ represents the metric induced on the worldsheet, while $h$ denotes its determinant. A useful quantity introduced in Ref.~\cite{Letelier:1979ej} is the antisymmetric bivector $\Sigma^{\mu\nu}$ associated with the worldsheet. It is defined by

\begin{equation}
\Sigma^{\mu\nu}
=
\epsilon^{AB}
\frac{\partial x^\mu}{\partial \xi^A}
\frac{\partial x^\nu}{\partial \xi^B},
\end{equation}
where $\epsilon^{AB}$ is the two-dimensional Levi--Civita symbol satisfying
$\epsilon^{01}=-\epsilon^{10}=-1$.
From the Nambu--Goto action one can derive the energy--momentum tensor of an individual string. For a continuous distribution, or cloud, of strings, Ref.~\cite{Letelier:1979ej} shows that the energy--momentum tensor can be written as

\begin{equation} \label{TensorEMNubeCuerdas}
    T^{\mu\nu}
    =
    -
    \frac{\rho_{cs}\,
    \Sigma^{\mu\alpha}
    \Sigma_{\alpha}^{\;\;\nu}}
    {\sqrt{-h}}
    \, ,
    \qquad
    h
    =
    \frac{1}{2}
    \Sigma_{\mu\nu}\Sigma^{\mu\nu}.
\end{equation}

For the configuration considered in Ref.~\cite{Letelier:1979ej}, the only nonvanishing components of the energy--momentum tensor satisfy
$T^0_{\;0}=T^1_{\;1}$.
Imposing the conservation law
$\nabla_\mu T^{\mu\nu}=0$
leads to

\begin{equation} \label{TensorEMNubeCuerdas1}
-T^0_0
=
-T^1_1
=
\rho
=
\sqrt{-h}\,\rho_{cs}
=
\frac{a}{r^2},
\end{equation}
where $\rho_{cs}$ is interpreted as the proper energy density of the string cloud. Consequently, the energy density of the cloud decays as $r^{-2}$, a characteristic feature that will be relevant for the discussion presented in the main text.

\section{A Brief Overview of the String-Fluid Model}

The cloud-of-strings model was generalized in Ref.~\cite{Letelier:1983du} by introducing an effective pressure, giving rise to the so-called string-fluid model. Within this framework, the energy--momentum tensor takes the form

\begin{equation}\label{TensorEMfluid}
T^{\mu\nu}
=
\left(
p+\rho_{cs}\sqrt{-h}
\right)
\frac{\Sigma^{\mu\lambda}\Sigma^{\nu}{}_{\lambda}}
{-h}
+
p\,g^{\mu\nu},
\end{equation}
where $p$ and $\rho_{cs}$ denote the pressure and the proper energy density of the string fluid, respectively. The bivector $\Sigma^{\mu\nu}$ and the determinant $h$ of the induced worldsheet metric were introduced in the previous appendix. For the static and spherically symmetric line element considered in the present work, Ref.~\cite{Soleng:1993yr} showed that the only nonvanishing components of $\Sigma^{\mu\nu}$ are $\Sigma^{tr}$ and $\Sigma^{\theta\phi}$, implying that $h<0$. Consequently, the corresponding energy--momentum tensor assumes the anisotropic form $T^\mu_{\ \nu}
=
\mathrm{diag}
\left(
-\rho,
-\rho,
p_t,
p_t
\right)$, so that the radial pressure satisfies $p_r=-\rho$, 
whereas the tangential pressure depends on the particular string-fluid configuration.

\bibliography{mybib.bib}

@article{Lin:2013ofa,
    author = "Lin, Kai and Li, Jin and Yang, Shuzheng",
    title = "{Quasinormal Modes of Hayward Regular Black Hole}",
    doi = "10.1007/s10773-013-1682-4",
    journal = "Int. J. Theor. Phys.",
    volume = "52",
    pages = "3771--3778",
    year = "2013"
}

@article{Iyer:1986np,
    author = "Iyer, Sai and Will, Clifford M.",
    title = "{Black Hole Normal Modes: A {WKB} Approach. 1. Foundations and Application of a Higher Order {WKB} Analysis of Potential Barrier Scattering}",
    reportNumber = "Print-86-1482 (WASH. U., ST. LOUIS)",
    doi = "10.1103/PhysRevD.35.3621",
    journal = "Phys. Rev. D",
    volume = "35",
    pages = "3621",
    year = "1987"
}

@Article{Konoplya:2003ii,
  author        = {Konoplya, R. A.},
  journal       = {Phys. Rev. D},
  title         = {{Quasinormal behavior of the d-dimensional Schwarzschild black hole and higher order WKB approach}},
  year          = {2003},
  pages         = {024018},
  volume        = {68},
  archiveprefix = {arXiv},
  doi           = {10.1103/PhysRevD.68.024018},
}

@Article{Konoplya:2019hlu,
  author        = {Konoplya, R. A. and Zhidenko, A. and Zinhailo, A. F.},
  journal       = {Class. Quant. Grav.},
  title         = {{Higher order WKB formula for quasinormal modes and grey-body factors: recipes for quick and accurate calculations}},
  year          = {2019},
  pages         = {155002},
  volume        = {36},
  archiveprefix = {arXiv},
  doi           = {10.1088/1361-6382/ab2e25},
  primaryclass  = {gr-qc},
}

@article{DeLorenzo:2014pta,
    author = "De Lorenzo, Tommaso and Pacilio, Costantino and Rovelli, Carlo and Speziale, Simone",
    title = "{On the Effective Metric of a Planck Star}",
    eprint = "1412.6015",
    archivePrefix = "arXiv",
    primaryClass = "gr-qc",
    doi = "10.1007/s10714-015-1882-8",
    journal = "Gen. Rel. Grav.",
    volume = "47",
    number = "4",
    pages = "41",
    year = "2015"
}

@article{Hayward:2005gi,
    author = "Hayward, Sean A.",
    title = "{Formation and evaporation of regular black holes}",
    eprint = "gr-qc/0506126",
    archivePrefix = "arXiv",
    doi = "10.1103/PhysRevLett.96.031103",
    journal = "Phys. Rev. Lett.",
    volume = "96",
    pages = "031103",
    year = "2006"
}

@article{Ovalle:2017fgl,
    author = "Ovalle, Jorge",
    title = "{Decoupling gravitational sources in general relativity: from perfect to anisotropic fluids}",
    eprint = "1704.05899",
    archivePrefix = "arXiv",
    primaryClass = "gr-qc",
    doi = "10.1103/PhysRevD.95.104019",
    journal = "Phys. Rev. D",
    volume = "95",
    number = "10",
    pages = "104019",
    year = "2017"
}

@article{Letelier:1979ej,
    author = "Letelier, P. S.",
    title = "{CLOUDS OF STRINGS IN GENERAL RELATIVITY}",
    doi = "10.1103/PhysRevD.20.1294",
    journal = "Phys. Rev. D",
    volume = "20",
    pages = "1294--1302",
    year = "1979"
}

@article{Halilsoy:2013iza,
    author = "Halilsoy, M. and Ovgun, A. and Mazharimousavi, S. Habib",
    title = "{Thin-shell wormholes from the regular Hayward black hole}",
    eprint = "1312.6665",
    archivePrefix = "arXiv",
    primaryClass = "gr-qc",
    doi = "10.1140/epjc/s10052-014-2796-4",
    journal = "Eur. Phys. J. C",
    volume = "74",
    pages = "2796",
    year = "2014"
}

@article{Bargueno:2020ais,
    author = "Bargue{\~n}o, Pedro",
    title = "{Some global, analytical and topological properties of regular black holes}",
    eprint = "2008.02680",
    archivePrefix = "arXiv",
    primaryClass = "gr-qc",
    doi = "10.1103/PhysRevD.102.104028",
    journal = "Phys. Rev. D",
    volume = "102",
    number = "10",
    pages = "104028",
    year = "2020"
}

@article{Bargueno:2021fus,
    author = "Bargue{\~n}o, Pedro",
    title = {{Singularity theorems for the Reissner-Nordstr{\"o}m spacetime and topology change in regular black holes}},
    doi = "10.1103/PhysRevD.104.024063",
    journal = "Phys. Rev. D",
    volume = "104",
    number = "2",
    pages = "024063",
    year = "2021"
}

@article{Melgarejo:2020mso,
    author = "Melgarejo, Gustavo and Contreras, Ernesto and Bargue{\~n}o, Pedro",
    title = "{Regular black holes with exotic topologies}",
    doi = "10.1016/j.dark.2020.100709",
    journal = "Phys. Dark Univ.",
    volume = "30",
    pages = "100709",
    year = "2020"
}

@article{Borde:1996df,
    author = "Borde, Arvind",
    title = "{Regular black holes and topology change}",
    eprint = "gr-qc/9612057",
    archivePrefix = "arXiv",
    doi = "10.1103/PhysRevD.55.7615",
    journal = "Phys. Rev. D",
    volume = "55",
    pages = "7615--7617",
    year = "1997"
}

@article{Letelier:1983du,
    author = "Letelier, Patricio S.",
    title = "{FLUIDS OF STRINGS IN GENERAL RELATIVITY}",
    reportNumber = "INIS-mf-7391",
    month = "7",
    year = "1983"
}

@article{Soleng:1993yr,
    author = "Soleng, Harald H.",
    title = "{Dark matter and nonNewtonian gravity from general relativity coupled to a fluid of strings}",
    eprint = "gr-qc/9412053",
    archivePrefix = "arXiv",
    reportNumber = "NORDITA-94-69",
    doi = "10.1007/BF02107935",
    journal = "Gen. Rel. Grav.",
    volume = "27",
    pages = "367--378",
    year = "1995"
}

@article{LIGOScientific:2016aoc,
    author = "Abbott, B. P. and others",
    collaboration = "LIGO Scientific, Virgo",
    title = "{Observation of Gravitational Waves from a Binary Black Hole Merger}",
    eprint = "1602.03837",
    archivePrefix = "arXiv",
    primaryClass = "gr-qc",
    reportNumber = "LIGO-P150914",
    doi = "10.1103/PhysRevLett.116.061102",
    journal = "Phys. Rev. Lett.",
    volume = "116",
    number = "6",
    pages = "061102",
    year = "2016"
}

@article{Zafar:2025sxl,
    author = "Zafar, Usman and Bamba, Kazuharu and Rasheed, Tabinda and Bhattacharya, Krishnakanta",
    title = "{Thermodynamic analysis of black holes with cloud of strings and quintessence via Barrow entropy}",
    eprint = "2504.00416",
    archivePrefix = "arXiv",
    primaryClass = "hep-th",
    reportNumber = "FU-PCG-151 FU-PCG-151 FU-PCG-151",
    doi = "10.1016/j.physletb.2025.139446",
    journal = "Phys. Lett. B",
    volume = "864",
    pages = "139446",
    year = "2025"
}

@article{Javed:2026tdz,
    author = {Javed, Faisal and Eid, A. and Bouzenada, Abdelmalek and Waseem, Arfa and Mustapha, N. and Akhmedov, Munisbek and Turaev, Yunus and G{\"u}dekli, Ertan},
    title = "{Phase structure, thermal fluctuations and emission spectrum of nonlinear electrodynamics AdS black holes surrounded by cloud of strings}",
    doi = "10.1016/j.dark.2025.102207",
    journal = "Phys. Dark Univ.",
    volume = "51",
    pages = "102207",
    year = "2026"
}

@article{Gogoi:2025ied,
    author = "Gogoi, Naba Jyoti and Gogoi, Dhruba Jyoti and Bora, Jyatsnasree",
    title = "{Topology of 5-dimensional Einstein{\textendash}Gauss{\textendash}Bonnet AdS black hole thermodynamics surrounded by a cloud of Strings}",
    doi = "10.1016/j.dark.2025.102099",
    journal = "Phys. Dark Univ.",
    volume = "50",
    pages = "102099",
    year = "2025"
}

@article{Ahmed:2025ojg,
    author = "Ahmed, Faizuddin and Bouzenada, Abdelmalek and Silva, Edilberto O.",
    title = "{AdS black hole solution with a dark matter halo surrounded by a cloud of strings}",
    eprint = "2509.10829",
    archivePrefix = "arXiv",
    primaryClass = "gr-qc",
    doi = "10.1140/epjc/s10052-025-15113-w",
    journal = "Eur. Phys. J. C",
    volume = "85",
    number = "12",
    pages = "1385",
    year = "2025"
}

@article{NunesdosSantos:2025alw,
    author = "Nunes dos Santos, Luis Cesar",
    title = "{Revisiting black holes surrounded by cloud and fluid of strings in general relativity}",
    eprint = "2502.15846",
    archivePrefix = "arXiv",
    primaryClass = "gr-qc",
    doi = "10.1103/PhysRevD.111.064032",
    journal = "Phys. Rev. D",
    volume = "111",
    number = "6",
    pages = "064032",
    year = "2025"
}

@article{Estrada:2026kea,
    author = "Estrada, Milko",
    title = "{Reissner Nordstrom black holes with integrable singularity interiors supported by string distributions}",
    eprint = "2602.03050",
    archivePrefix = "arXiv",
    primaryClass = "gr-qc",
    doi = "10.1140/epjc/s10052-026-15888-6",
    journal = "Eur. Phys. J. C",
    volume = "86",
    pages = "774",
    year = "2026"
}

@article{Huo:2026enq,
    author = "Huo, Tianxu and Liu, Chengzhou",
    title = "{Perihelion precession and gyroscopic geodesic precession in Schwarzschild spacetime with a fluid of strings background}",
    doi = "10.1016/j.dark.2026.102252",
    journal = "Phys. Dark Univ.",
    volume = "52",
    pages = "102252",
    year = "2026"
}

@article{Errehymy:2026ntw,
    author = "Errehymy, A. and Turimov, B. and Khan, M. A. and Usanov, S. and Yasakov, Z. and Avezmuratova, Z.",
    title = "{Slowly rotating traversable wormholes supported by radially varying string-fluid matter: From regular geometries to photon trajectories}",
    eprint = "2606.11261",
    archivePrefix = "arXiv",
    primaryClass = "gr-qc",
    month = "6",
    year = "2026"
}

@article{Estrada:2024lhk,
    author = "Estrada, Milko and Crispim, Tiago Mota and Alencar, Geova",
    title = "{A Way of Decoupling the Gravitational Bulk Field Equations~of Regular Braneworld Black Holes to Suppress the Bulk Singularities}",
    eprint = "2410.06189",
    archivePrefix = "arXiv",
    primaryClass = "gr-qc",
    doi = "10.1002/prop.202400220",
    journal = "Fortsch. Phys.",
    volume = "73",
    number = "3",
    pages = "2400220",
    year = "2025"
}

@article{Guimaraes:2025jsh,
    author = "Guimar{\~a}es, V. F. and Cavalcanti, R. T. and da Rocha, R.",
    title = "{Hair imprints of the gravitational decoupling and hairy black hole spectroscopy}",
    eprint = "2506.20044",
    archivePrefix = "arXiv",
    primaryClass = "gr-qc",
    doi = "10.1088/1361-6382/adfda5",
    journal = "Class. Quant. Grav.",
    volume = "42",
    number = "17",
    pages = "175011",
    year = "2025"
}

@article{Sallah:2026dhk,
    author = "Sallah, Malick and Moneer, Eman M. and Sharif, M. and Zotos, Euaggelos E.",
    title = "{From black holes to wormholes via minimal geometric deformation in rastall theory}",
    doi = "10.1016/j.physletb.2026.140171",
    journal = "Phys. Lett. B",
    volume = "873",
    pages = "140171",
    year = "2026"
}

@article{Alshammari:2026wnf,
    author = "Alshammari, Mohammad and Rizwan, M. and Almatroud, Othman Abdullah and Bhatti, M. Z. and Alshammari, Saleh and Yousaf, Z.",
    title = "{Imprints of holographic dark energy and minimally deformed wormholes in general relativity}",
    doi = "10.1016/j.dark.2026.102219",
    journal = "Phys. Dark Univ.",
    volume = "51",
    pages = "102219",
    year = "2026"
}

@article{Hua:2025qwu,
    author = "Hua, Yaobin and Ban, Zhenglong and Ren, Tian-You and Yin, Jia-Jun and Yang, Rong-Jia",
    title = "{Regular hairy black holes through gravitational decoupling method}",
    eprint = "2510.20524",
    archivePrefix = "arXiv",
    primaryClass = "gr-qc",
    doi = "10.1140/epjc/s10052-026-15287-x",
    journal = "Eur. Phys. J. C",
    volume = "86",
    number = "1",
    pages = "44",
    year = "2026"
}

@article{Naseer:2025ghn,
    author = "Naseer, Tayyab and Levi Said, Jackson and Sharif, M. and Abdel-Aty, Abdel-Haleem",
    title = "{Thermodynamic properties of non-singular Hayward black hole through the lens of minimal gravitational decoupling}",
    doi = "10.1140/epjc/s10052-025-14186-x",
    journal = "Eur. Phys. J. C",
    volume = "85",
    number = "4",
    pages = "471",
    year = "2025"
}

@article{Lukash:2013ts,
    author = "Lukash, Vladimir N. and Strokov, Vladimir N.",
    title = "{Space-Times with Integrable Singularity}",
    eprint = "1301.5544",
    archivePrefix = "arXiv",
    primaryClass = "gr-qc",
    doi = "10.1142/S0217751X13500073",
    journal = "Int. J. Mod. Phys. A",
    volume = "28",
    pages = "1350007",
    year = "2013"
}

@article{Estrada:2023dcj,
    author = "Estrada, Milko and Alencar, G. and Crispim, Tiago M. and Muniz, C. R.",
    title = "{New models of d-dimensional black holes without inner horizon and with an integrable singularity}",
    eprint = "2310.01734",
    archivePrefix = "arXiv",
    primaryClass = "gr-qc",
    doi = "10.1088/1475-7516/2024/06/042",
    journal = "JCAP",
    volume = "06",
    pages = "042",
    year = "2024"
}

@article{Estrada:2024moz,
    author = "Estrada, Milko",
    title = "{A new representation of vacuum Lovelock solutions in d=2N+1 dimensions: Black holes with an integrable singularity and regular black holes}",
    eprint = "2411.01253",
    archivePrefix = "arXiv",
    primaryClass = "gr-qc",
    doi = "10.1016/j.aop.2025.169985",
    journal = "Annals Phys.",
    volume = "477",
    pages = "169985",
    year = "2025"
}

@article{Casadio:2026tmd,
    author = "Casadio, Roberto and Giusti, Andrea and Kamenshchik, Alexander and Ovalle, Jorge",
    title = "{On gravitational collapse and integrable singularities}",
    eprint = "2605.01808",
    archivePrefix = "arXiv",
    primaryClass = "gr-qc",
    doi = "10.1140/epjc/s10052-026-16258-y",
    journal = "Eur. Phys. J. C",
    volume = "86",
    number = "8",
    pages = "1018",
    year = "2026"
}

@article{Ovalle:2023vvu,
    author = "Ovalle, Jorge",
    title = "{Black holes without Cauchy horizons and integrable singularities}",
    eprint = "2305.00030",
    archivePrefix = "arXiv",
    primaryClass = "gr-qc",
    doi = "10.1103/PhysRevD.107.104005",
    journal = "Phys. Rev. D",
    volume = "107",
    number = "10",
    pages = "104005",
    year = "2023"
}

@article{Carballo-Rubio:2022pzu,
    author = "Carballo-Rubio, Ra{\'u}l and Di Filippo, Francesco and Liberati, Stefano and Pacilio, Costantino and Visser, Matt",
    title = "{Comment on {\textquotedblleft}Regular evaporating black holes with stable cores{\textquotedblright}}",
    eprint = "2212.07458",
    archivePrefix = "arXiv",
    primaryClass = "gr-qc",
    reportNumber = "YITP-22-154",
    doi = "10.1103/PhysRevD.108.128501",
    journal = "Phys. Rev. D",
    volume = "108",
    number = "12",
    pages = "128501",
    year = "2023"
}

@article{Bonanno:2022jjp,
    author = "Bonanno, Alfio and Khosravi, Amir-Pouyan and Saueressig, Frank",
    title = "{Regular evaporating black holes with stable cores}",
    eprint = "2209.10612",
    archivePrefix = "arXiv",
    primaryClass = "gr-qc",
    doi = "10.1103/PhysRevD.107.024005",
    journal = "Phys. Rev. D",
    volume = "107",
    number = "2",
    pages = "024005",
    year = "2023"
}

@article{Bonanno:2025bgc,
    author = "Bonanno, Alfio and Panassiti, Antonio and Saueressig, Frank",
    title = "{Cauchy horizon stability and instability of regular black holes}",
    eprint = "2507.03581",
    archivePrefix = "arXiv",
    primaryClass = "gr-qc",
    doi = "10.1103/34w6-s2yl",
    journal = "Phys. Rev. D",
    volume = "114",
    number = "6",
    pages = "064027",
    year = "2026"
}

@article{Ghosh:2022gka,
    author = "Ghosh, Rajes and Rahman, Mostafizur and Mishra, Akash K.",
    title = "{Regularized stable Kerr black hole: cosmic censorships, shadow and quasi-normal modes}",
    eprint = "2209.12291",
    archivePrefix = "arXiv",
    primaryClass = "gr-qc",
    doi = "10.1140/epjc/s10052-023-11252-0",
    journal = "Eur. Phys. J. C",
    volume = "83",
    number = "1",
    pages = "91",
    year = "2023"
}

@article{Ma:2014qma,
    author = "Ma, Meng-Sen and Zhao, Ren",
    title = "{Corrected form of the first law of thermodynamics for regular black holes}",
    eprint = "1411.0833",
    archivePrefix = "arXiv",
    primaryClass = "gr-qc",
    doi = "10.1088/0264-9381/31/24/245014",
    journal = "Class. Quant. Grav.",
    volume = "31",
    pages = "245014",
    year = "2014"
}

@article{Estrada:2020tbz,
    author = "Estrada, Milko and Tello-Ortiz, Francisco",
    title = "{A new model of regular black hole in (2+1) dimensions}",
    eprint = "2012.05068",
    archivePrefix = "arXiv",
    primaryClass = "gr-qc",
    doi = "10.1209/0295-5075/ac0ed0",
    journal = "EPL",
    volume = "135",
    number = "2",
    pages = "20001",
    year = "2021"
}

@article{Estrada:2023cyx,
    author = "Estrada, Milko and Aros, Rodrigo",
    title = "{A new class of regular black holes in Einstein Gauss-Bonnet gravity with localized sources of matter}",
    eprint = "2305.17233",
    archivePrefix = "arXiv",
    primaryClass = "gr-qc",
    doi = "10.1016/j.physletb.2023.138090",
    journal = "Phys. Lett. B",
    volume = "844",
    pages = "138090",
    year = "2023"
}

@article{Estrada:2026sze,
    author = "Estrada, Milko",
    title = "{Integrable Singularity Black Holes without Inner Horizons in Asymptotically Safe Inspired Gravity and NFW Dark Matter Effects}",
    eprint = "2611.15813",
    archivePrefix = "arXiv",
    primaryClass = "gr-qc",
    month = "11",
    year = "2026"
}

@article{Ashtekar:2011ni,
    author = "Ashtekar, Abhay and Singh, Parampreet",
    title = "{Loop Quantum Cosmology: A Status Report}",
    eprint = "1108.0893",
    archivePrefix = "arXiv",
    primaryClass = "gr-qc",
    doi = "10.1088/0264-9381/28/21/213001",
    journal = "Class. Quant. Grav.",
    volume = "28",
    pages = "213001",
    year = "2011"
}

@article{Kelly:2020uwj,
    author = "Kelly, Jarod George and Santacruz, Robert and Wilson-Ewing, Edward",
    title = "{Effective loop quantum gravity framework for vacuum spherically symmetric spacetimes}",
    eprint = "2006.09302",
    archivePrefix = "arXiv",
    primaryClass = "gr-qc",
    doi = "10.1103/PhysRevD.102.106024",
    journal = "Phys. Rev. D",
    volume = "102",
    number = "10",
    pages = "106024",
    year = "2020"
}

@article{Husain:2006cx,
    author = "Husain, Viqar and Winkler, Oliver",
    title = "{Quantum Hamiltonian for gravitational collapse}",
    eprint = "gr-qc/0601082",
    archivePrefix = "arXiv",
    doi = "10.1103/PhysRevD.73.124007",
    journal = "Phys. Rev. D",
    volume = "73",
    pages = "124007",
    year = "2006"
}

@article{Rovelli:2014cta,
    author = "Rovelli, Carlo and Vidotto, Francesca",
    title = "{Planck stars}",
    eprint = "1401.6562",
    archivePrefix = "arXiv",
    primaryClass = "gr-qc",
    doi = "10.1142/S0218271814420267",
    journal = "Int. J. Mod. Phys. D",
    volume = "23",
    number = "12",
    pages = "1442026",
    year = "2014"
}

@article{Rovelli:2017zoa,
    author = "Rovelli, Carlo",
    title = "{Planck stars as observational probes of quantum gravity}",
    eprint = "1708.01789",
    archivePrefix = "arXiv",
    primaryClass = "gr-qc",
    doi = "10.1038/s41550-017-0065",
    journal = "Nature Astron.",
    volume = "1",
    pages = "0065",
    year = "2017"
}

@article{Lewandowski:2022zce,
    author = "Lewandowski, Jerzy and Ma, Yongge and Yang, Jinsong and Zhang, Cong",
    title = "{Quantum Oppenheimer-Snyder and Swiss Cheese Models}",
    eprint = "2210.02253",
    archivePrefix = "arXiv",
    primaryClass = "gr-qc",
    doi = "10.1103/PhysRevLett.130.101501",
    journal = "Phys. Rev. Lett.",
    volume = "130",
    number = "10",
    pages = "101501",
    year = "2023"
}

@article{Adler:2001vs,
    author = "Adler, Ronald J. and Chen, Pisin and Santiago, David I.",
    title = "{The Generalized uncertainty principle and black hole remnants}",
    eprint = "gr-qc/0106080",
    archivePrefix = "arXiv",
    reportNumber = "SLAC-PUB-8853",
    doi = "10.1023/A:1015281430411",
    journal = "Gen. Rel. Grav.",
    volume = "33",
    pages = "2101--2108",
    year = "2001"
}

@article{Estrada:2025lgo,
    author = "Estrada, Milko",
    title = "{Black hole solutions with a linear equation of state in Ho{\v{r}}ava gravity and Einstein{\textendash}{\ae}ther theory}",
    eprint = "2512.20445",
    archivePrefix = "arXiv",
    primaryClass = "gr-qc",
    doi = "10.1016/j.physletb.2026.140798",
    journal = "Phys. Lett. B",
    volume = "880",
    pages = "140798",
    year = "2026"
}

\end{document}